\documentclass[twocolumn]{aastex631}

\usepackage{amsmath}
\usepackage[symbol]{footmisc}

\begin{document}

\title{Comparing Globular Cluster System Properties with Host Galaxy Environment\footnote{Based on observations made with the NASA/ESA Hubble Space Telescope, obtained from the Data Archive at the Space Telescope Science Institute, which is operated by the Association of Universities for Research in Astronomy, Inc., under NASA contract NAS5-26555.  These observations are associated with programs 14219 and 15265.}}

\author{Kate Hartman}
\affiliation{Department of Physics \& Astronomy, McMaster University, 1280 Main St W, Hamilton, ON L8S 1T7, Canada}

\author{William E. Harris}
\affiliation{Department of Physics \& Astronomy, McMaster University, 1280 Main St W, Hamilton, ON L8S 1T7, Canada}

\author{John P. Blakeslee}
\affiliation{NOIRLab, 950 N. Cherry Ave., Tucson, AZ 85719, USA}

\author{Chung-Pei Ma}
\affiliation{Department of Astronomy and Department of Physics, University of California, Berkeley, CA 94720, USA}

\author{Jenny E. Greene}
\affiliation{Department of Astrophysical Sciences, Princeton University, Princeton, NJ 08544, USA}

\begin{abstract}

We present \textit{Hubble Space Telescope} photometry in optical (F475X) and near-infrared (F110W) bands of the globular cluster (GC) systems of the inner halos of a sample of 15 massive elliptical galaxies.  The targets are selected from the volume-limited MASSIVE survey, and chosen to sample a range of environments from sparsely populated groups to BCGs in dense clusters.  We also present a quantitative model of the relation between (F475X - F110W) colour and cluster metallicity [M/H], using simulated GCs.   Because much of the GC population in such galaxies is built up through accretion, the metallicity distribution of the GC systems might be expected to vary with galaxy environment.  The photometry is used to create a completeness-corrected metallicity distribution for each galaxy in the sample, and to fit a double Gaussian curve to each histogram in order to model the two standard red and blue subpopulations.  Finally, the properties of the GC metallicity distribution are correlated against galaxy environment.  We find that almost no GCS properties and host galaxy environmental properties are correlated, with the exception of a weak but consistent correlation between blue fraction and \textit{n}th-nearest neighbour surface density.  The results suggest that the systemic properties of the GCS, at least in the inner to mid-halo regions, are influenced more strongly by the local environment at early times, rather than by the environmental properties we see today.

\end{abstract}



\section{Introduction} \label{sec:intro}

Globular clusters (GCs) are old, massive, dense, gravitationally bound systems of stars.  They are nearly ubiquitous in galaxies, found in all but the least massive of dwarfs \citep[e.g.][]{harris2010review,forbes2018metgrad,beasley2020review}.  As some of the earliest stellar structures to form within their host galaxies, they are powerful probes of the early history of hierarchical growth and chemical enrichment.

The most populous globular cluster systems (GCSs) belong to massive elliptical (early-type or ETG) galaxies.  Several GCS properties scale with those of the host galaxy, perhaps most notably total GCS mass and galaxy halo mass \citep{blakeslee1997scaling,spitler2009scaling,harris2013stellarmass,hudson2014scaling}.  Massive galaxies typically have distinguishable subpopulations of blue, metal-poor GCs and red, metal-rich GCs \citep[e.g.][]{brodie2006colour,usher2012colour,brodie2012colour}---although this is not universally the case, with occasional examples of unimodal or multimodal populations known \citep[see e.g.][]{blom2012ngc4365,harris2016sigmoid,harris2017BCGs,beasley2018unimodal}. 

Simulations such as in \cite{pillepich2018simgrowth} and \cite{elbadry2019assembly} support a hierarchical merger model of galaxy growth in which massive elliptical galaxies grow to their present sizes by merging with smaller satellite galaxies, and sometimes other large galaxies; other simulation and model work such as E-MOSAICS \citep[][etc.]{pfeffer2018emosaics,kruijssen2019emosaics,bastian2020emosaics,horta2021scaling,reinacampos2020hierarchical}. \cite{kravtsov2005hierarchical}, \cite{boylankolchin2017scaling}, and \cite{choksi2018scaling} have been able to reproduce GC scaling relations in a hierarchical merger framework.  This process leads to a wide range of evolutionary histories, which differ depending on a galaxy's mass and location in its environment.  In rich galaxy clusters, the most massive and luminous elliptical galaxy will settle near the center, and these brightest cluster galaxies (BCGs) draw in satellite galaxies from their surroundings, growing their own stellar and halo masses and adding the satellites' GCs to their own GCSs (cf.~the references cited above).  Meanwhile, galaxies in low-density areas have access to less accretable material than a BCG in a rich galaxy cluster.  It is natural to ask whether or not the present-day GCS properties of BCGs depend on environment.

\cite{choksi2019scaling} addressed this problem in a coarse way (see their Figure 4) by comparing full merger trees (including accreted satellites) to the main progenitor branch only (excluding accretion), finding that only the full set of accretions matched the observations then available (they used observations of the Virgo Cluster from \cite{peng2008comps}).  Observers are beginning to tackle this problem as well; studies such as \cite{cho2012isolated}, \cite{salinas2015isolated}, \cite{beasley2018unimodal}, and \cite{ennis2019isolated} specifically targeted isolated elliptical galaxies rather than BCGs in rich clusters, although the subject galaxies in those studies are less massive and less luminous than those in our sample, and \cite{debortoli2022scaling} focused on environment as a driver of difference between GCSs, although that work looked at GCS structure rather than GCS metallicity.

In this work, we explore the unaddressed parameter space of GCS metallicity versus host galaxy environment.  Section \ref{sec:data} presents our data and our photometry techniques, Section \ref{sec:completeness} outlines our procedure to account for incompleteness, and Section \ref{sec:metallicity} explains our conversions from GC color to metallicity.  In Section \ref{sec:DGfits}, we characterize the shape of each galaxy's GCS metallicity distribution function, and Section \ref{sec:environment} compares GCS parameters to galaxy environment metrics.  We discuss our findings and summarize our work in Section \ref{sec:summary}.

\section{Data and photometry} \label{sec:data}

The strong scaling relations between GCS properties and host galaxy stellar and halo mass put constraints on how the sample of galaxies for this work could be constructed.  In an unrestricted sample of galaxies, the relations involving mass drown out more subtle second-order signals from relations such as those involving environment \citep{harris2013stellarmass,debortoli2022scaling}.  Our sample had to be constructed to minimize differences in stellar and halo mass, and therefore to minimize differences in GCS properties related to galaxy mass.
Our sample comprises fifteen galaxies from the MASSIVE survey sample \citep{ma2014massive}.  MASSIVE targeted the most luminous galaxies within $\sim 100$ Mpc
and with stellar masses $M_* > 10^{11.5} M_{\odot}$.  Our selected targets all lie  within a very narrow stellar mass range, and thus effectively controlled for that variable.

\subsection{Images} \label{subsec:images}

\begin{figure*}
    \figurenum{1}
    \includegraphics[width=0.93\textwidth]{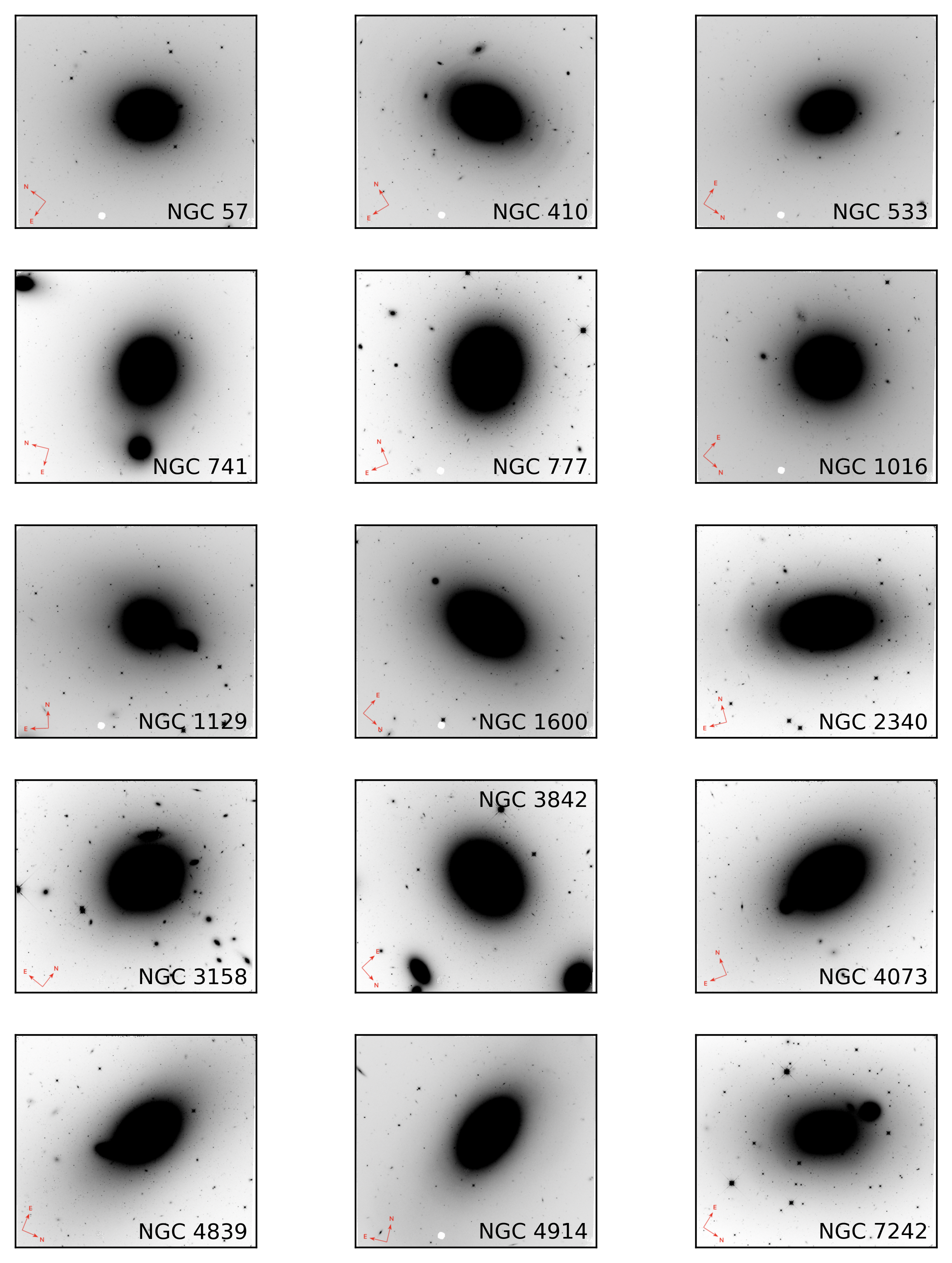}
    \caption{F110W (1.1 $\mu$m) images of all fifteen galaxies in our sample from \textit{HST} GO programs 14219 and 15265.}
    \label{fig:f110w_images}
\end{figure*}

This work makes use of \textit{HST} archival data from previous work targeting the MASSIVE galaxies \citep[][GO proposal 14219]{goullaud2018f110w}) along with more recent images from GO proposal 15265.  The images were taken with \textit{HST}'s WFC3 instrument using the F475X (475 nm) and F110W (1.1 $\mu$m) filters.  The total exposure time for each image was approximately one orbit.  See Table \ref{tab:basics} for observing information and Figure \ref{fig:f110w_images} for the F110W images.  The data is available at MAST: \dataset[10.17909/pvve-1002]{\doi{10.17909/pvve-1002}}.

\begin{table*}
    \centering
    \caption{Basic observing and physical data}
    \label{tab:basics}
    \scriptsize
    \begin{tabular}{c|ccccccccccc} 
        \toprule
        Galaxy & GO number & \shortstack{F475X \\ exposure \\ time (s)} & \shortstack{F110W \\ exposure \\ time (s)} & \shortstack{Distance \\ (Mpc)\tablenotemark{\tiny a}} & $M_V$\tablenotemark{\tiny d} & $M_K$\tablenotemark{\tiny e} & $E_{F475X}$\tablenotemark{\tiny g} & $E_{F110W}$\tablenotemark{\tiny g} & $M_*$ ($\log M_{\odot}$)\tablenotemark{\tiny h} & $R_e$ (kpc)\tablenotemark{\tiny i} \\
        \tableline
        NGC 57 & 15265, 14219 & 2394 & 2496 & 73 & -22.8 & -25.75 & 0.241 & 0.068 & 11.79 & 9.52 \\
        NGC 410 & 15265, 14219 & 2421 & 2496 & 71 & -23.1 & -25.90 & 0.183 & 0.052 & 11.86 & 10.95 \\
        NGC 533 & 15265, 14219 & 2382 & 2496 & 75 & -22.8 & -26.05 & 0.095 & 0.027 & 11.92 & 14.77 \\
        NGC 741 & 15265, 14219 & 2385 & 2496 & 75 & -23.1 & -26.06 & 0.164 & 0.046 & 11.93 & 9.84 \\
        NGC 777 & 15265, 14219 & 2421 & 2496 & 68 & -22.8 & -25.94 & 0.145 & 0.041 & 11.87 & 6.13 \\
        NGC 1016 & 15265, 14219 & 2382 & 2496 & 92 & -22.9 & -26.33 & 0.097 & 0.027 & 12.05 & 11.93 \\
        NGC 1129 & 15265, 14219 & 2478 & 2496 & 72 & -21.9 & -26.14 & 0.352 & 0.099 & 11.96 & 10.47 \\
        NGC 1600 & 15265, 14219 & 2385 & 2496 & 66 & -22.6 & -25.99 & 0.134 & 0.038 & 11.90 & 6.65 \\
        NGC 2340 & 15265 & 2658 & 2812 & 86 & -23.1 & -25.90 & 0.231 & 0.065 & 11.86 & 8.20 \\
        NGC 3158 & 15265 & 2538 & 2612 & 103 & -23.3 & -26.28 & 0.041 & 0.012 & 12.02 & 8.07 \\
        NGC 3842 & 15265 & 2496 & 2612 & 95 & -22.9 & -25.91 & 0.067 & 0.019 & 11.86 & 11.13 \\
        NGC 4073 & 15265 & 2484 & 2612 & 89\tablenotemark{\tiny b} & -23.3 & -26.33 & 0.084 & 0.024 & 12.05 & 9.92 \\
        NGC 4839 & 15265 & 2505 & 2612 & 110\tablenotemark{\tiny c} & -23.2 & -25.85 & 0.031 & 0.009 & 11.83 & 15.57 \\
        NGC 4914 & 15265, 14219 & 2439 & 2496 & 70 & -22.7 & -25.72 & 0.042 & 0.012 & 11.78 & 10.61 \\
        NGC 7242 & 15265 & 2538 & 2612 & 78 & -21.5 & -26.34 & 0.472 & 0.133 & 12.05 & 10.51 \\
        \tableline
    \end{tabular}
    \tablenotetext{}{(a) calculated from 
    \cite{vandenbosch2015distanceref} unless otherwise indicated; (b) calculated from \cite{sdss2017distanceref}; (c) calculated from \cite{rines2016distanceref}; (d) \cite{deVauc1991catalogue}; (e) \cite{ma2014massive}; (f) \cite{crook2007groups}; (g) \cite{schlafly2011reddening}; (h) \cite{veale2017massive}; (i) \cite{sdss2011dr8}}
\end{table*}

\subsection{Environmental data: galaxies} \label{subsec:galaxies}

The MASSIVE survey selection criteria included a K-band luminosity cutoff of $M_K < -25.3$, which corresponded to a stellar mass cutoff of $M_* \gtrsim 10^{11.5}M_{\odot}$ \citep{ma2014massive}.  MASSIVE galaxies also have relatively high Galactic latitudes and low foreground reddenings, so the observed fields have few contaminating field stars.  Galaxy group and cluster memberships are given in Table \ref{tab:basics}.  Distances listed in the Table assume $H_0 = 70$ km s$^{-1}$ Mpc$^{-1}$ and mean CMB frame radial velocities from \cite{vandenbosch2015distanceref}, \cite{sdss2017distanceref}, and \cite{rines2016distanceref}.  It should be noted that these distances are systematically slightly greater than the more recent surface brightness fluctuation distances of \cite{jensen2021distances} \citep[see also][]{blakeslee2021distances}, but only by $\sim 4$ Mpc in most cases.  These slight offsets do not affect any of the later conclusions in this study.
The K-band magnitude and stellar mass data from \cite{ma2014massive} and \cite{veale2017massive}, also seen in Table \ref{tab:basics}, allowed us to test how well we had controlled for stellar mass in our galaxy sample and to ensure that GCS-galaxy mass relations would not overpower any GCS-environment relations.

\subsection{Environmental data: groups} \label{subsec:groups}

\begin{figure*}
    \figurenum{2}
    \centering
    \includegraphics[width=\textwidth]{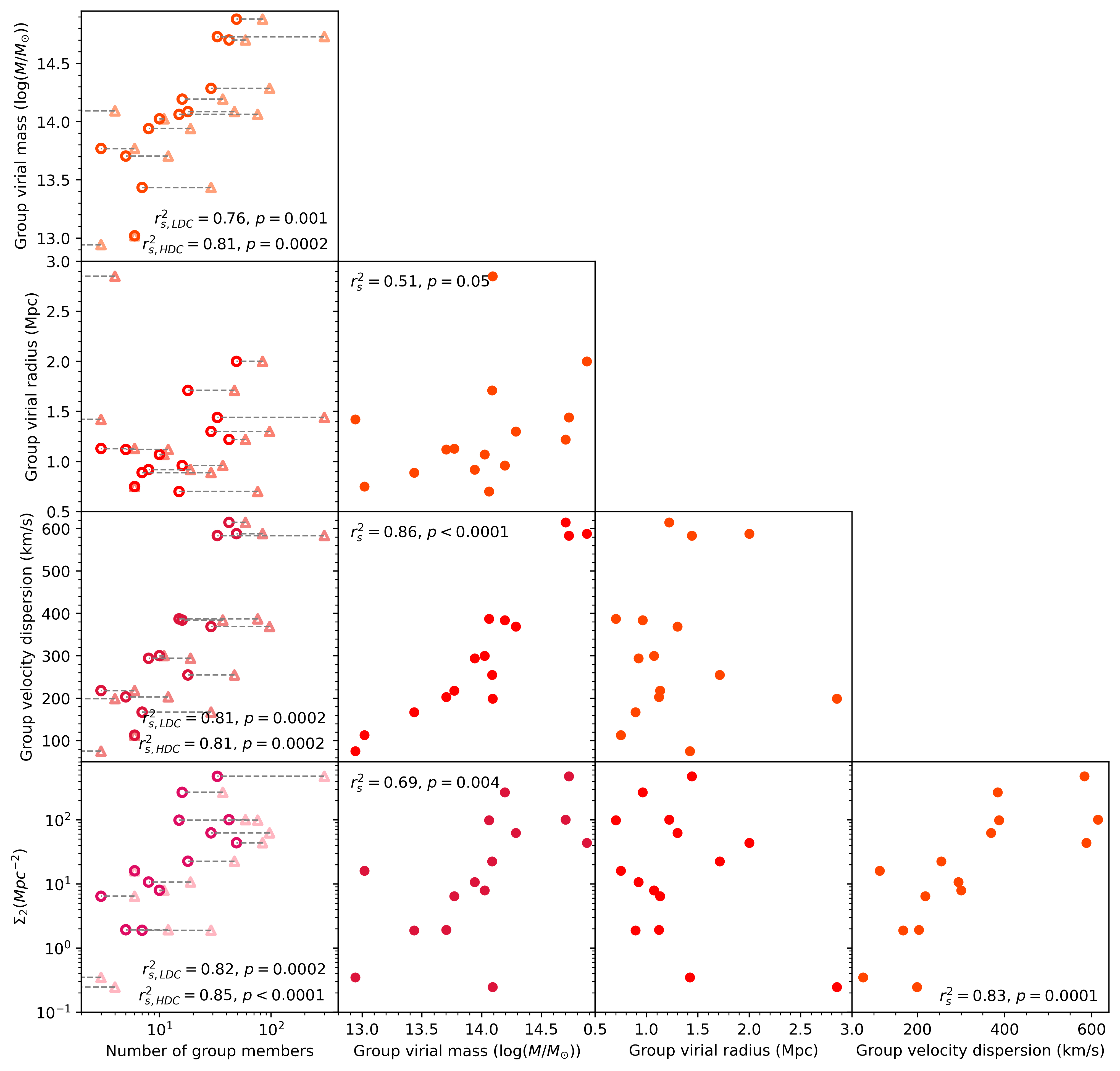}
    \caption{Cross-checks of environmental variables, with Spearman correlation coefficients if the correlation is significant.  All data except $\Sigma_n$ is taken directly from \cite{crook2007groups}; $\Sigma_{n}$ is derived from \cite{crook2007groups} coordinates and distances from Table \ref{tab:basics}.  In the leftmost column, open circles indicate HDC measurements, and open triangles LDC measurements, with dashed lines connecting points for the same galaxy.}
    \label{fig:enviro_checks}
\end{figure*}

\begin{table*}
    \centering
    \caption{Environmental data}
    \label{tab:chm_data}
    \begin{tabular}{c|ccccccc} 
        \toprule
        Galaxy & \shortstack{Group \\ number}\tablenotemark{*} & \shortstack{Group \\ members}\tablenotemark{*} & \shortstack{Group virial \\ mass ($\log M_{\odot}$)}\tablenotemark{*} & \shortstack{Group virial \\ radius (Mpc)}\tablenotemark{*} & \shortstack{Group velocity \\ dispersion (km/s)}\tablenotemark{*} & $\Sigma_2$ (Mpc$^{-2})$ & $\Sigma_5$ (Mpc$^{-2})$ \\
        \tableline
        NGC 57 & L11 & 4 & 14.092 & 2.85 & 198.8 & 0.2453 & N/A \\
        NGC 410 & H53 & 29 & 14.286 & 1.30 & 368.8 & 62.41 & 5.000 \\
        NGC 533 & H74 & 3 & 13.769 & 1.13 & 217.9 & 6.416 & 0.6980 \\
        NGC 741 & H102 & 5 & 13.704 & 1.12 & 203.2 & 1.924 & 0.7893 \\
        NGC 777 & H109 & 7 & 13.434 & 0.89 & 167.2 & 1.891 & 0.4923 \\
        NGC 1016 & H159 & 8 & 13.940 & 0.92 & 294.1 & 10.69 & 11.38 \\
        NGC 1129 & H185 & 33 & 14.730 & 1.44 & 583.5 & 475.5 & 23.74 \\
        NGC 1600 & H294 & 16 & 14.193 & 0.96 & 384.2 & 267.7 & 7.348 \\
        NGC 2340 & H426 & 18 & 14.086 & 1.71 & 254.8 & 22.56 & 0.4929 \\
        NGC 3158 & H570 & 6 & 13.019 & 0.75 & 112.9 & 16.00 & 0.8733 \\
        NGC 3842 & H672 & 42 & 14.701 & 1.22 & 614.5 & 100.1 & 95.05 \\
        NGC 4073 & H692 & 10 & 14.024 & 1.07 & 300.3 & 7.943 & 7.631 \\
        NGC 4839 & H745 & 49 & 14.880 & 2.00 & 588.0 & 43.77 & 1.500 \\
        NGC 4914 & L929 & 3 & 12.943 & 1.42 & 75.1 & 0.3473 & N/A \\
        NGC 7242 & H1186 & 15 & 14.062 & 0.70 & 387.0 & 98.56 & 56.66 \\
        \tableline
    \end{tabular}
    \tablenotetext{*}{\cite{crook2007groups}}
\end{table*}

The galaxy environmental data used in this work were derived from the group catalogues of \cite{crook2007groups}.  Crook and collaborators created two catalogues of galaxy groups: a low density contrast (LDC) catalogue with less stringent group inclusion criteria, and a high density contrast (HDC) catalogue with more stringent criteria.  Thirteen of the fifteen galaxies in our sample appear in both the HDC and LDC, while the remaining two, NGC 57 and NGC 4914, are in very sparse groups and appear only in the LDC.  This work used HDC data when available, and LDC data only for NGC 57 and NGC 4914.  It should be noted that NGC 57's group virial radius from \cite{crook2007groups} is quite large compared to those for the rest of our sample; because of this and the more relaxed LDC inclusion criteria, we should be cautious when working with group data for NGC 57 and NGC 4914.

To characterize the density of each host galaxy's environment more directly, we used the coordinates provided by \cite{crook2007groups} to make two $n$th-nearest-neighbor measurements in projection, defined as in \cite{cooper2005neighbours}:

\begin{equation}
    \Sigma_n = \frac{n}{\pi D^2_{p,n}}
\end{equation} where $D_{p,n}$ is the projected distance to the $n$th galaxy from the galaxy of interest.  We made a calculation for part of the galaxy sample, excluding NGC 57 and NGC 4914, with $n=5$, a standard compromise value of \textit{n} that allows for both measurements of small galaxy groups and avoidance of ultra-small number statistics, and then a measurement for the whole sample with $n=2$, motivated by the NGC 4914 group, the smallest in the sample with three galaxies including NGC 4914 itself.

Figure \ref{fig:enviro_checks} compares all Crook-derived metrics, including $n$th-nearest-neighbor surface density for $n=2$.  Group mass $M$ goes as group velocity dispersion $\sigma^2$ and group radius $R$, so positive correlations between number of group members, group virial mass, and group velocity dispersion are expected given that the central BCGs in this sample all have similar stellar masses.  Spearman correlation coefficients are given in the Figure for all significant relations.

\subsection{Photometry} \label{subsec:photometry}

The galaxies in our sample are distant enough that their GCs appear as nearly unresolved point sources \citep{harris2017BCGs}.   At a distance of $d = 80$ Mpc, the 6-pc half-light diameter of a typical GC is equivalent to an angular width of $0.015''$, well below the $0.1''$ optical resolution of \textit{HST}.  With that in mind, we used DOLPHOT \citep{dolphin2000dolphot}, a program optimized for stellar photometry with \textit{HST} data, to measure our images.  DOLPHOT identified and measured the integrated light from each GC as it would for a star.

\begin{figure}
    \figurenum{3}
    \centering
    \includegraphics[width=0.45\textwidth]{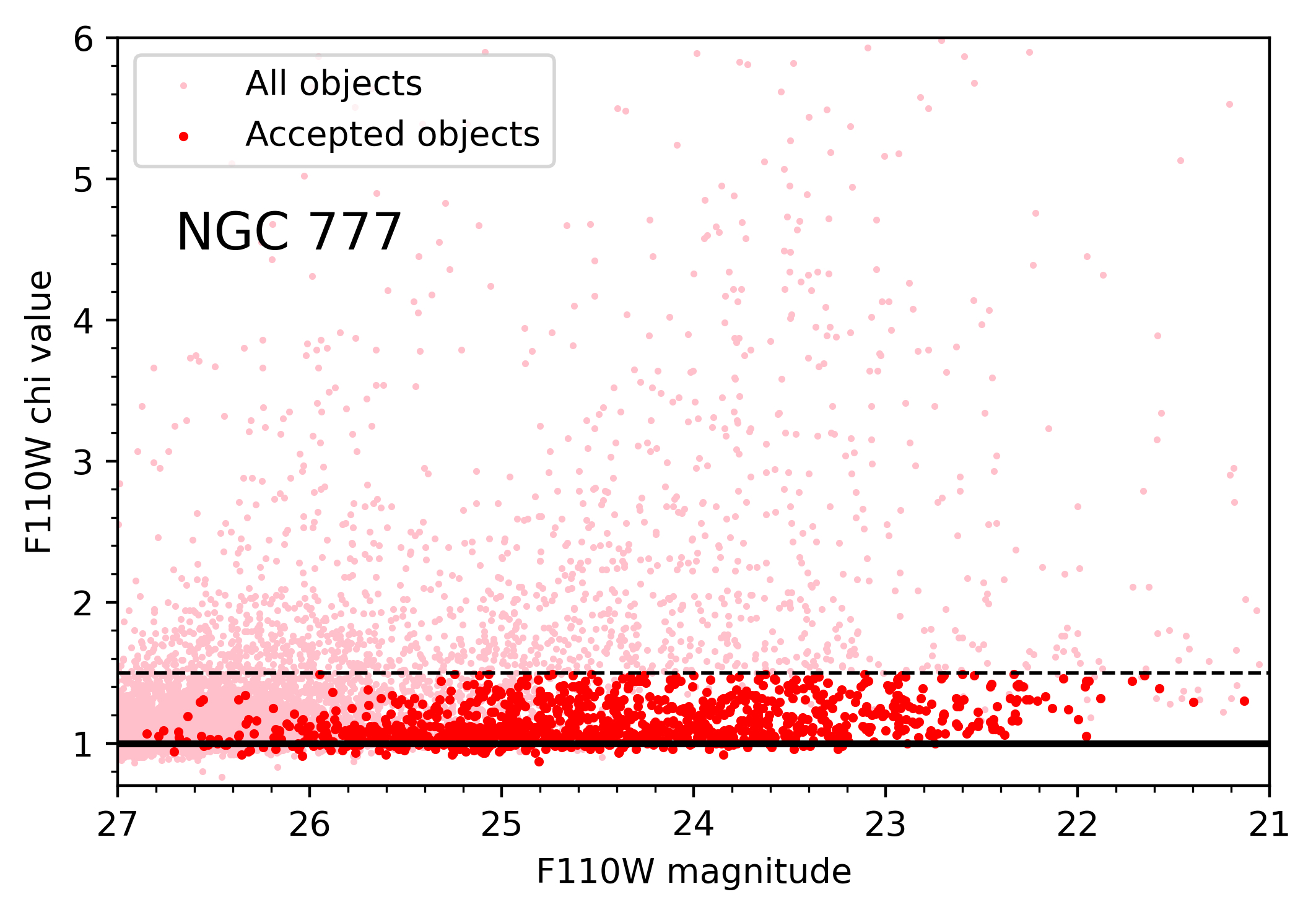}
    \caption{Chi vs. magnitude in F110W, for NGC 777.  This is a typical chi-magnitude distribution for our sample.  Small pink points denote all objected detected by DOLPHOT, and larger red dots denote objects remaining in the cleaned dataset.  The black dashed line shows the upper chi limit.}
    \label{fig:ngc777_chi}
\end{figure}

\begin{figure}
    \figurenum{4}
    \centering
    \includegraphics[width=0.45\textwidth]{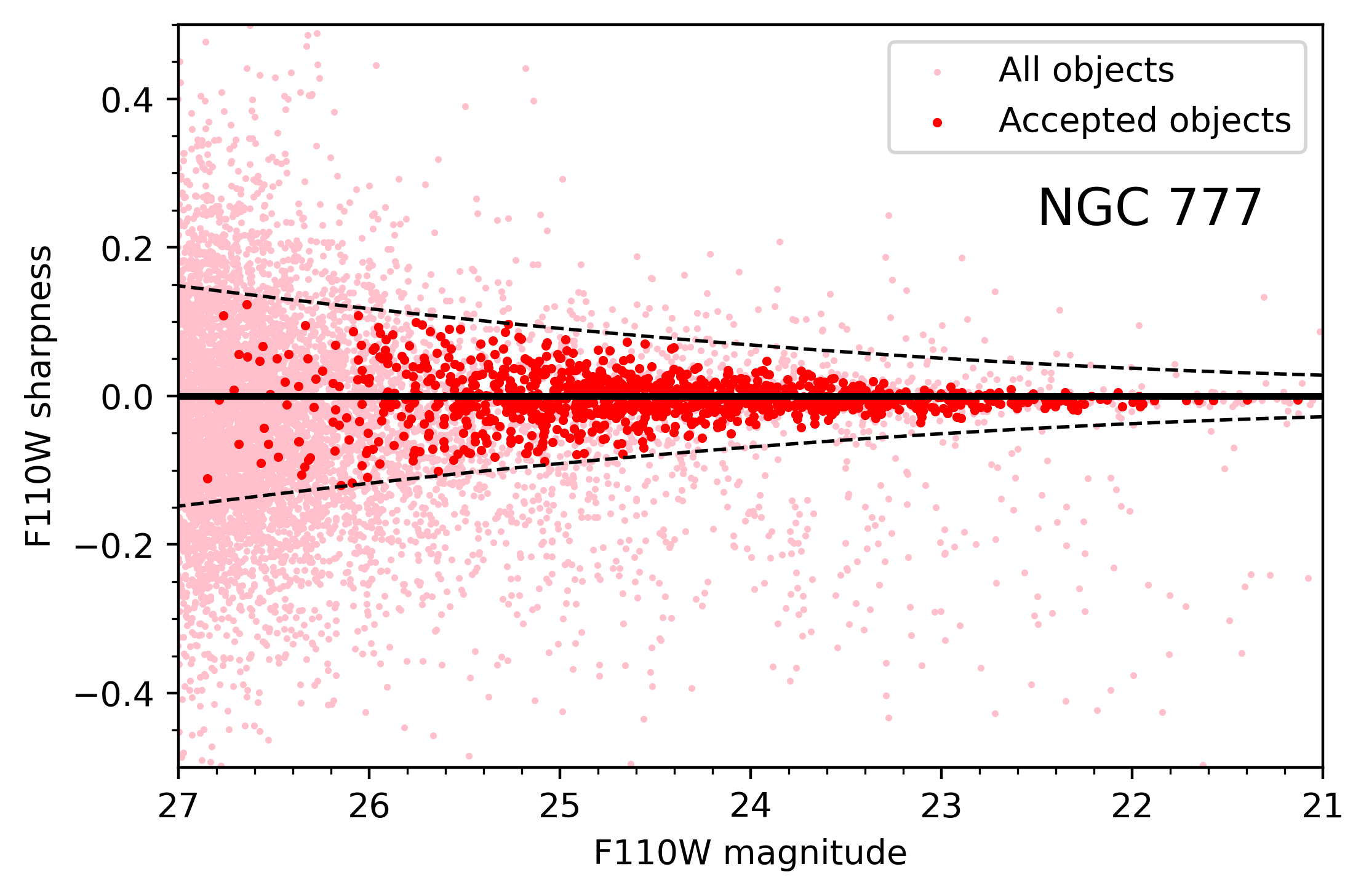}
    \caption{Sharpness vs. magnitude in F110W, for NGC 777.  This is a typical sharp-magnitude distribution for our sample.  Small pink points denote all objected detected by DOLPHOT, and larger red dots denote objects remaining in the cleaned dataset.  The black dashed line shows the upper and lower sharp limits.}
    \label{fig:ngc777_sharp}
\end{figure}

In addition to magnitudes on the VEGAMAG scale, DOLPHOT includes in its output several measurement quality flags and metrics for each detected object, including an object type flag (type 1 objects are point sources; other type numbers denote extended sources, blended sources, cosmic ray strikes, and objects too faint to measure), signal to noise ratio (SNR), chi (for measuring goodness of fit to the  standard point spread functions), and sharpness (a measure of object width relative to standard PSF width).  After running DOLPHOT, we cleaned the list of detected objects by retaining only those that met the following criteria:

\begin{itemize}
    \item DOLPHOT object type 1
    \item Magnitude $< 90$ in both filters, to reject objects too faint to be measured
    \item SNR $\geq 4$, to ensure high-quality magnitude measurements
    \item Chi $\leq 1.5$ in both filters, to ensure high-quality magnitude measurements (see Figure \ref{fig:ngc777_chi} for an example)
    \item $-0.15 <$ F475X sharpness $< 0.08$, to capture point sources (see Figure \ref{fig:ngc777_sharp} for an example)
    \item $|\textrm{F110W sharpness}| < 2.3(0.3608 - 0.0363(\textrm{F110W}) + 0.000938(\textrm{F110W})^2$, to capture point sources with room for scatter at the faint end
\end{itemize}

\begin{figure*}
    \figurenum{5}
    \centering
    \includegraphics[width=\textwidth]{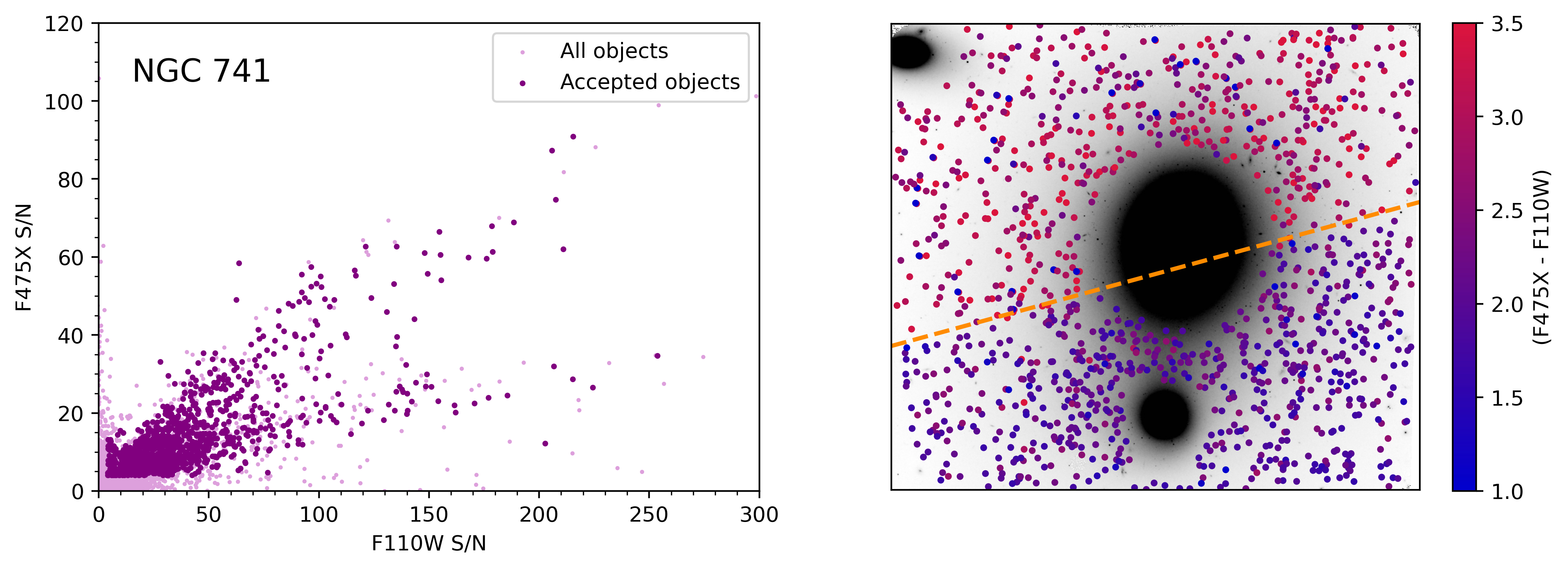}
    \caption{Data from NGC 741.  \textit{Left:} SNR for F110W vs. F475X.  There are two visible populations of objects, one with F475X SNR consistent with the rest of our data, and the other with lower F475X SNR.  \textit{Right:} an overlay of detected objects on an image of NGC 741.  The boundary between the two F475X chips falls on a diagonal through the center of the galaxy, denoted by the dashed orange line; objects in the top half of the image, with overly faint F475X magnitudes, are significantly redder than those in the bottom of the image.  It is unclear what happened to the data from the top chip; the bottom chip appears to be unaffected.}
    \label{fig:ngc741}
\end{figure*}

The sharpness and chi criteria are especially effective at removing nonstellar objects (small, faint background galaxies; camera artifacts; bad pixels; etc.) from the sample.  The remaining culled lists are completely dominated by the rich GC systems around these giant galaxies.  The F475X sharpness criterion is skewed slightly toward negative values (i.e. objects broader than the PSF) because WFC3 with the F475X filter has a higher resolution ($0.04''$ per pixel) than its IR channel used with F110W ($0.1''$ per pixel), so some GCs are are expected to be just marginally resolved in F475X \citep[see][]{harris2009metgrad}.  With regard to the last two criteria, we found the difference between the linear sharpness cut used for F475X and the parabolic cut used for F110W to be negligible.  Both choices were motivated by the actual distribution of sharpness measurements.

Close preliminary inspection of the photometry after all the culling steps listed above revealed an anomaly for one of the fields, NGC 741.  The SNR values indicated a significant number of bad measurements in F475X for this galaxy.  Upon further inspection, we found that one of the two CCD chips used with F475X had produced unrealistically faint magnitudes (see Figure \ref{fig:ngc741}) and thus extremely red colors on one side of the galaxy.  In our analysis, we have therefore used only objects from the other, apparently unaffected chip that had produced F475X measurements in line with those from the rest of our galaxy sample.  For this galaxy, the GC sample size is therefore roughly a factor of two smaller than originally intended.

\section{Completeness and consistency corrections} \label{sec:completeness}

\begin{figure*}
    \figurenum{6}
    \centering
    \includegraphics[width=\textwidth]{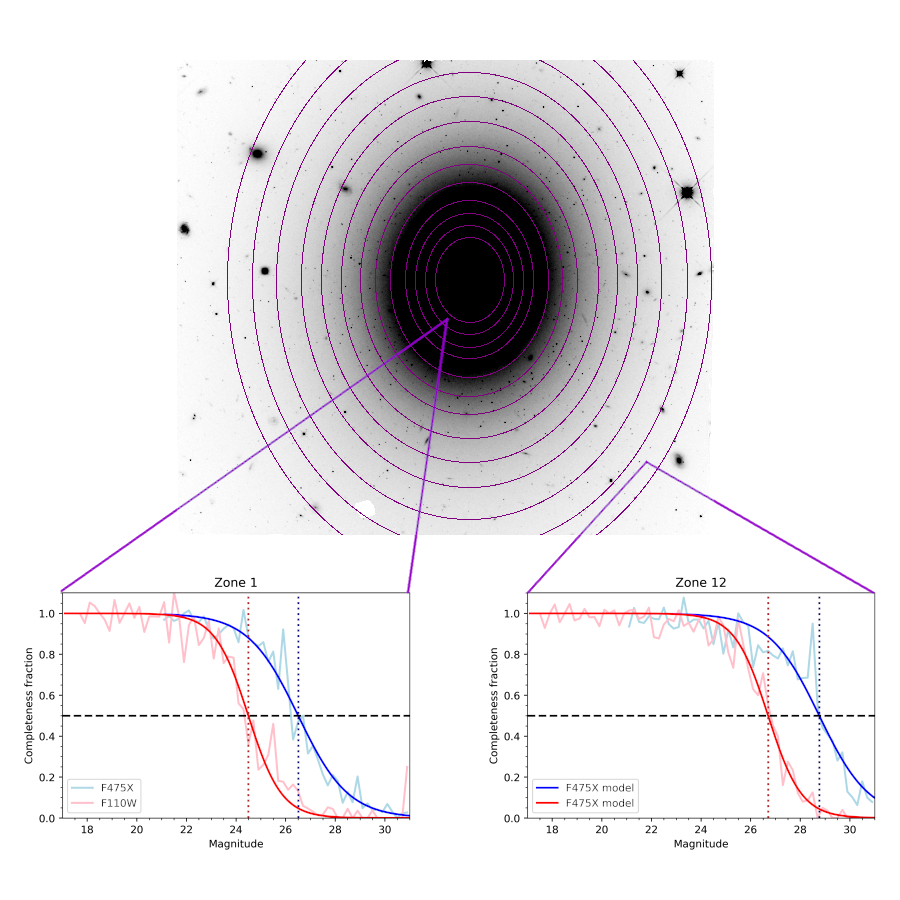}
    \caption{Elliptical zones for the NGC 777 artificial star test, with sigmoids for the innermost (1) and outermost (12) zones.  Pale lines trace the recovered star fraction, and darker lines are sigmoid models, with F475X shown in blue and F110W in red.  The horizontal dashed line denotes the 50\% completeness level, and the vertical dashed lines mark $m_0$ for each filter.}
    \label{fig:zones}
\end{figure*}

Our images posed a challenge when it came to characterizing the detection completeness of our GC samples: the background light from the host galaxy has a strong radial gradient from galaxy center outward (particularly in F110W), so the resulting sky noise therefore obscures more GCs toward the center of each image than at the outskirts.  In order to accurately count GCs in all areas of our images, we used the DOLPHOT artificial-star test function to perform an extensive completeness study on three of the fields.  We created twelve radial zones centered on each galaxy---circular zones for NGC 1016, and elliptical zones of constant eccentricity for NGC 57 and NGC 777 as seen in the top of Figure \ref{fig:zones}.  We added 3500 artificial stars to each zone, and determined the artificial star recovery fraction for each zone as a function of magnitude.  We then modeled the recovery fraction in each zone as a sigmoid curve \citep{harris2016sigmoid}:

\begin{equation}
    f_{comp} = \frac{1}{1+e^{\alpha(m-m_0)}}
    \label{eq:sigmoid}
\end{equation}

\begin{figure}
    \figurenum{7}
    \centering
    \includegraphics[width=0.47\textwidth]{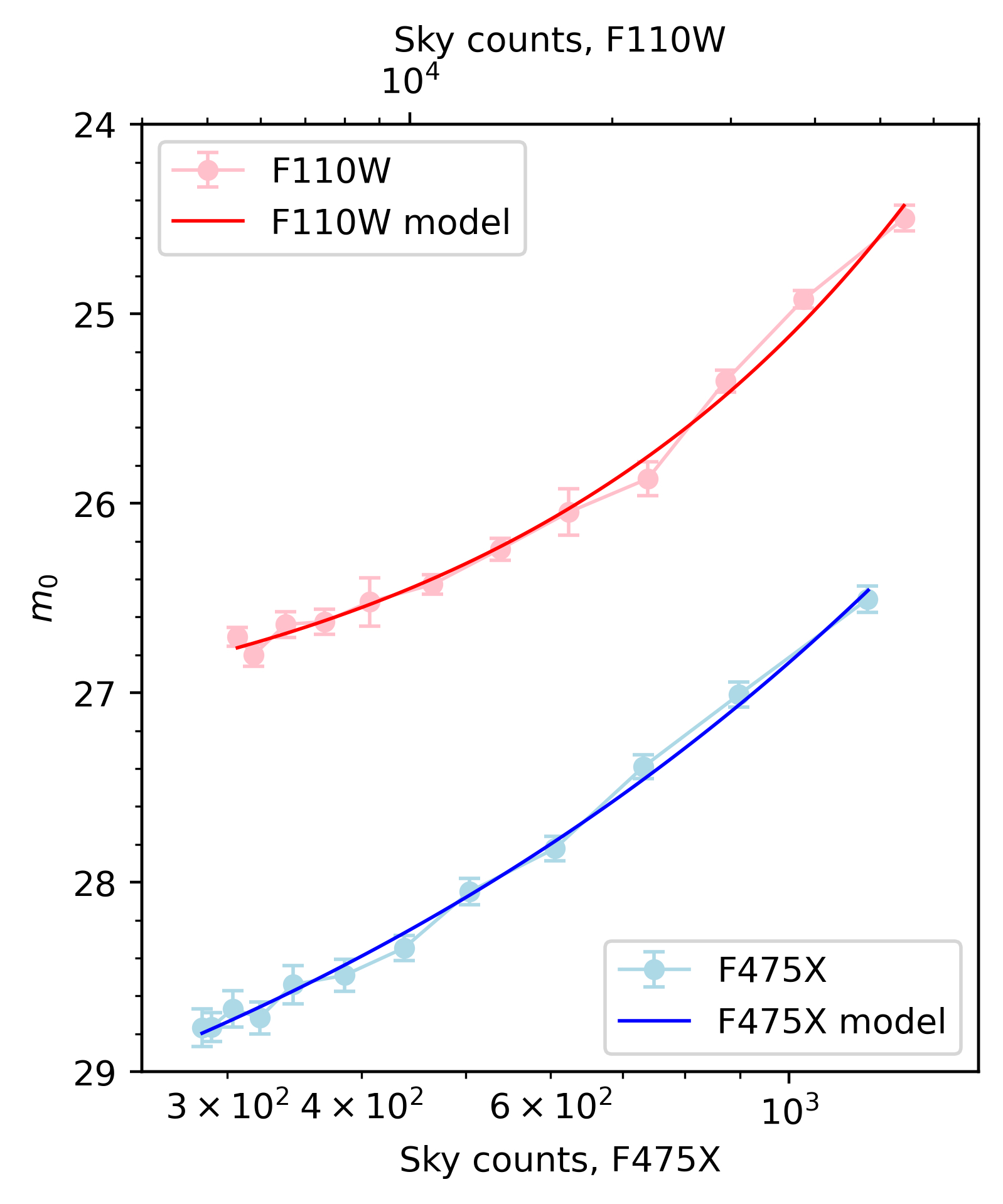}
    \caption{50\% completion magnitude $m_0$ vs. sky brightness from NGC 777 for F475X (blue) and F110W (pink) with fitted exponential curves.}
    \label{fig:m_0_sky}
\end{figure}

\begin{figure}
\figurenum{8}
    \centering
    \includegraphics[width=0.47\textwidth]{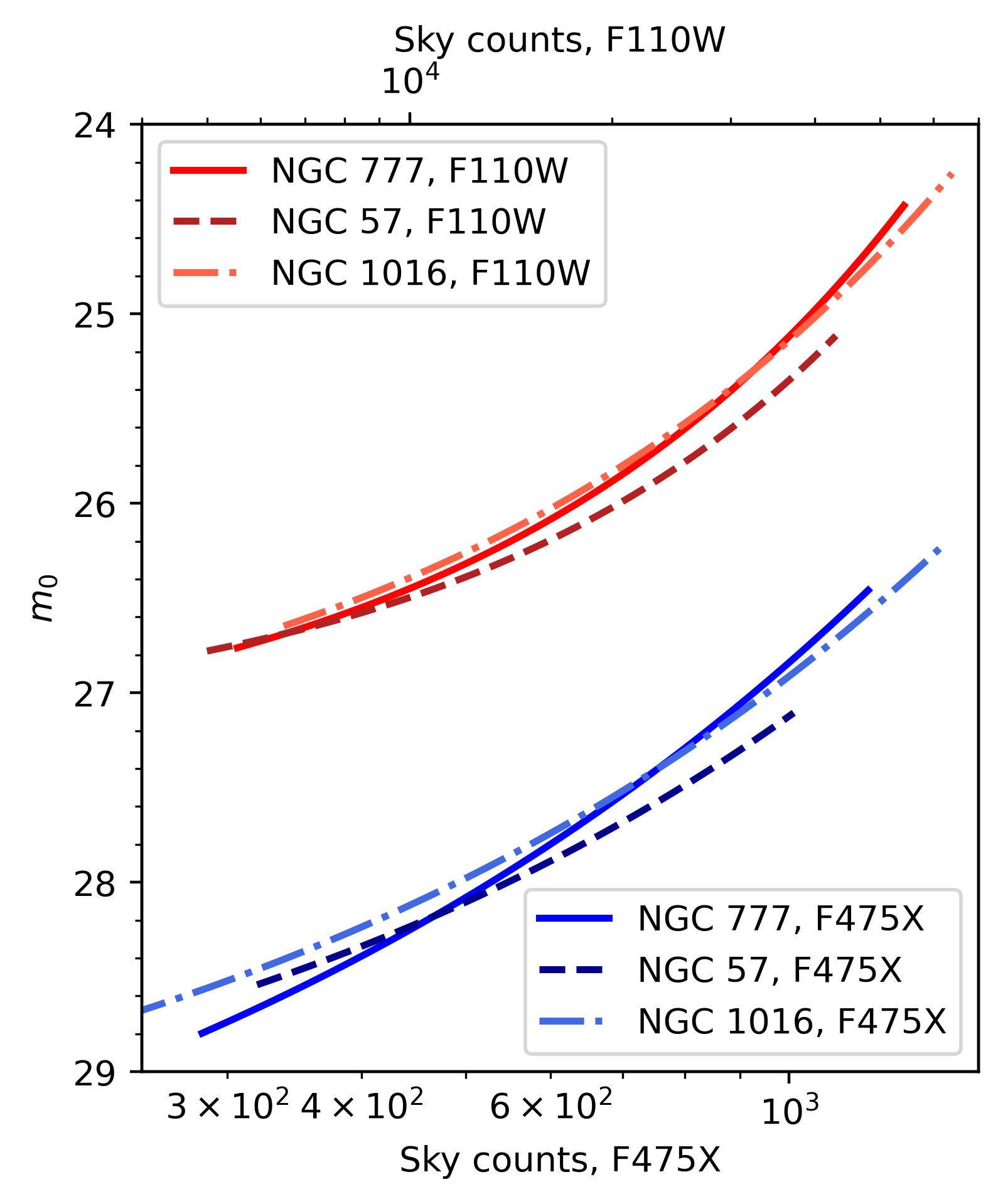}
    \caption{Exponential fits for $m_0$ vs. sky brightness in F475X (blue) and F110W (red) showing agreement between all three artificial star tests.}
    \label{fig:all_m_0_sky}
\end{figure}

Eq. \ref{eq:sigmoid} accounts for the slope $\alpha$ of the transition region where objects quickly become too faint to be detected and the magnitude $m_0$ at which half of the existing objects are detected, and produces a completeness fraction $f_{comp}$, which can be interpreted as a probability: how likely are we to detect an object of a given magnitude?  Because of the gradient of sky background level, $m_0$ is itself a function of location on the image, changing smoothly with galactocentric radius (see the bottom of Figure \ref{fig:zones}, comparing 50\% completeness levels in the innermost and outermost zones around NGC 777).  After determining a constant $\alpha$ for both filters, we found $m_0$ for each of the twelve zones and compared it to background brightness.  The relation is well fitted with an exponential curve (see Figure \ref{fig:m_0_sky} for the results from NGC 777).  We found that all three of our test galaxies produced similar relations in both filters (Figure \ref{fig:all_m_0_sky}), as expected since the filters and exposure times were essentially identical for all the fields, so we adopted the average exponential parameters for $m_0$ as functions of local sky brightness $\beta$ for our entire sample of galaxies:

\begin{equation}
    m_{0,\textrm{F}475\textrm{X}} = -0.05497\beta_{\rm F475X}^{0.6024} + 30.17
    \label{eq:m_0_f475x}
\end{equation}

\begin{equation}
    m_{0,\textrm{F}110\textrm{W}} = -0.003423\beta_{\rm F110W}^{0.6570} + 27.37
    \label{eq:m_0_f110w}
\end{equation}

This sky brightness-dependent completeness relation allowed us to more accurately characterize completeness for GCs throughout our images, regardless of how close they were to the centers of their host galaxies.  In order to ensure that our samples of GCs would accurately represent each GCS, we calculated $f_{comp}$ for each object from its magnitude and local sky brightness, and removed all objects from the dataset with a completeness fraction less than 0.5 in either filter.

Finally, to ensure that comparisons between the galaxies would be made based on identical ranges in both GC luminosity and radial region of the halo, we made an absolute magnitude cut based on the faintest object in NGC 4839 (the most distant galaxy in the sample), and a radial distance cut at $\sim20$ kpc based on NGC 1600 (the nearest galaxy in the sample).  These cuts ensured that we were sampling approximately the same portion of each GCS.  

\begin{figure*}
    \figurenum{9}
    \includegraphics[width=\textwidth]{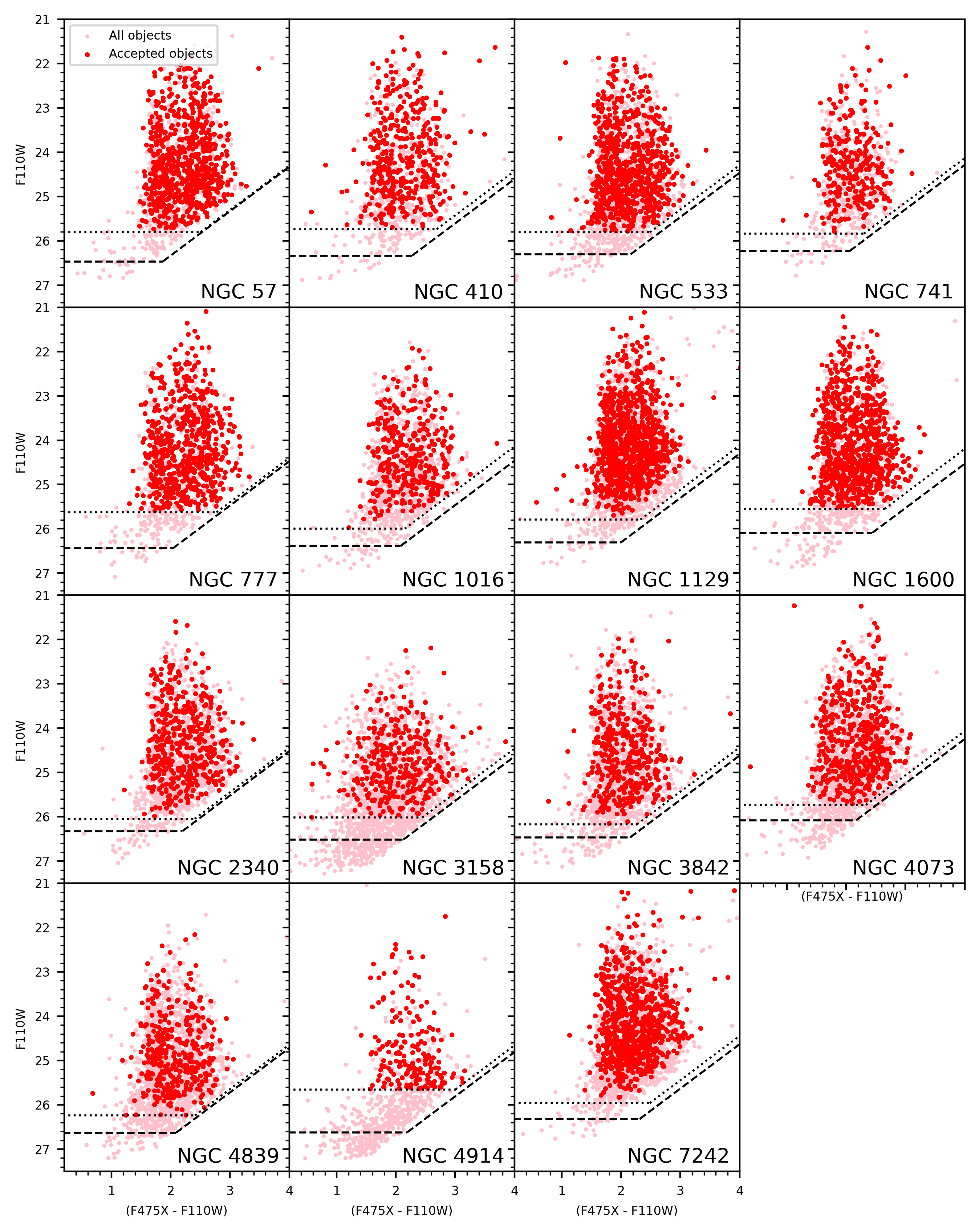}
    \caption{CMDs for our sample of galaxies.  Small pink points denote all detected objects; larger red points denote the final cleaned dataset.  The dashed lines mark 50\% completeness limits based on the completeness correction procedure, and the dotted lines mark effective 50\% completeness limits imposed by the radial and magnitude cutoffs.}
    \label{fig:f110w_cmds}
\end{figure*}

The results of our data cleanup procedure can be seen in Figure \ref{fig:f110w_cmds}.  All objects detected by DOLPHOT are shown as small pink points, the 50\% completeness limits as determined by our artificial star tests are denoted by the dashed lines, and the effective completeness limits imposed by the final absolute-magnitude and radial cuts are denoted by the dotted lines.  All objects shown as bigger red points are included in the final datasets for each galaxy that are used to define the color and metallicity distributions for its GCS.

\section{Color-metallicity conversion} \label{sec:metallicity}

Although integrated GC color is an observationally efficient proxy for metallicity, the relationship between color and metallicity is monotonic but nonlinear to some degree for optical/NIR colors \citep[e.g.][among others]{brodie2006colour,peng2006nonlinear,harris2010review,usher2015nonlinear,fahrion2020spectroscopy}, though \cite{villaume2019spectroscopy} models the relation for $(g-z)$ linearly. To draw conclusions involving GC or GCS composition, it is necessary to adopt a conversion of the GCS color distribution functions (CDFs) to metallicity distribution functions (MDFs).

Because spectroscopic metallicity measurements of extragalactic GCs require major observing campaigns \citep[e.g.][]{villaume2019spectroscopy,fahrion2020spectroscopy}, most color indices have not been calibrated this way, including the index used in this work, ($\textrm{F}475\textrm{X} - \textrm{F}110\textrm{W}$).  Instead of converting directly from color to metallicity, we used the methodology described in \citet{harris2023}, performing a two-step process using a different, spectroscopically calibrated \textit{HST} color index along with simulated GCs built with SSP (single stellar population) models.  The stellar models adopted here are the widely used Osservatorio Astronomico di Padova suite in its CMD 3.6 version (\href{http://stev.oapd.inaf.it/cgi-bin/cmd_3.6}{online tool}).

The HST color index with the strongest currently available spectroscopic calibration is ($\textrm{F}475\textrm{W} - \textrm{F}850\textrm{LP}$), equivalent to $(g-z)$ in the SDSS system \citep{peng2006nonlinear,usher2015nonlinear,sinnott2010colour,villaume2019spectroscopy,fahrion2020spectroscopy}.  To quantify \textit{HST}'s ($\textrm{F}475\textrm{W} - \textrm{F}850\textrm{LP}$) VEGAMAG index versus metallicity, we used a simple quadratic relation to the spectroscopic data as described more completely in \citet{harris2023}, which is essentially a combination of the transformations in the spectroscopic studies cited above:

\begin{multline}
    (\textrm{F}475\textrm{W} - \textrm{F}850\textrm{LP}) = 2.158 \\ + 0.57081[\textrm{Fe}/\textrm{H}] + 0.10026[\textrm{Fe}/\textrm{H}]^2
    \label{eq:spectroscopy}
\end{multline}

\begin{figure}
    \figurenum{10}
    \centering
    \includegraphics[width=0.47\textwidth]{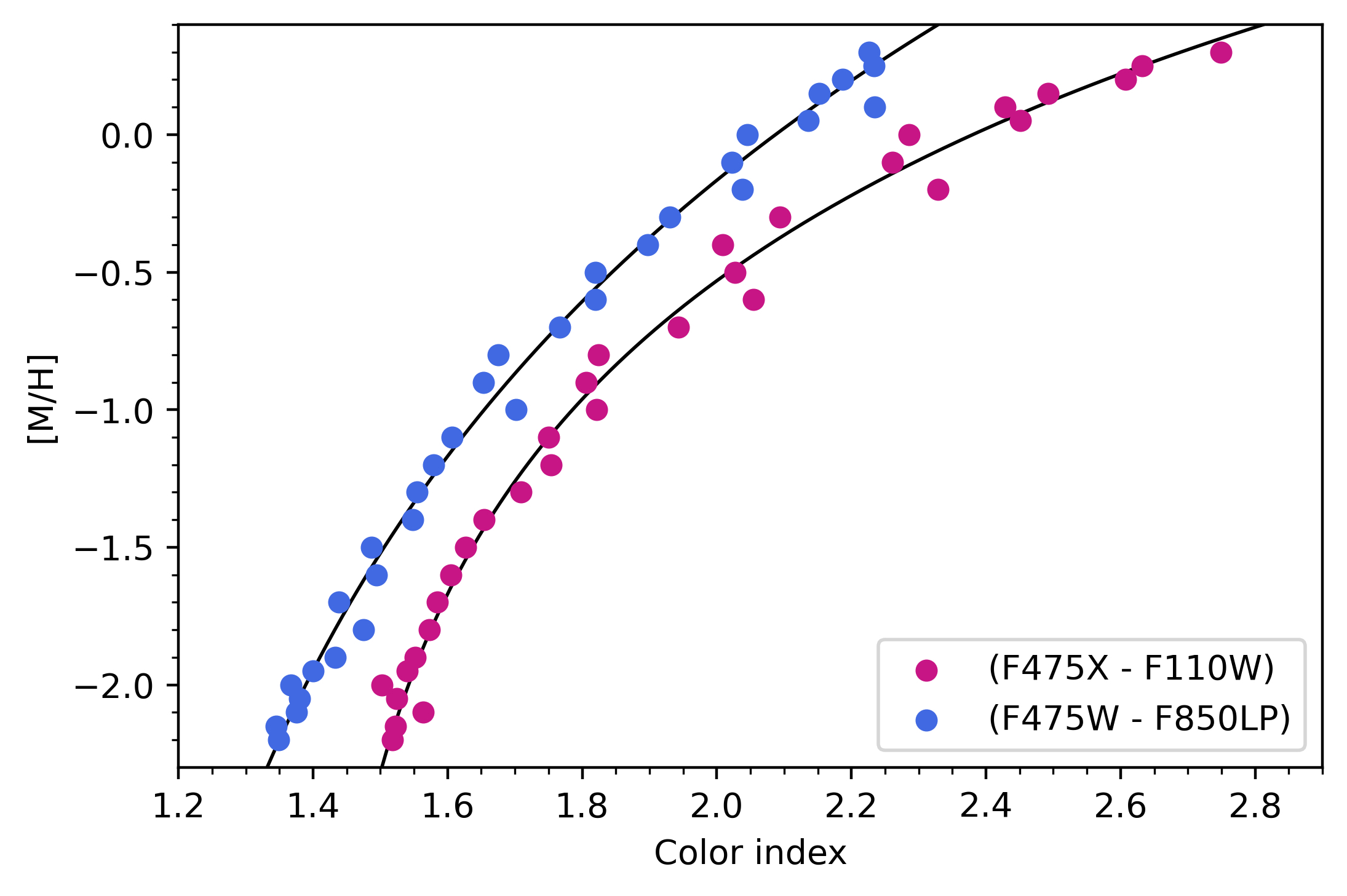}
    \caption{Sets of simulated observations of GCs of known metallicity from the Padova online tool, in (F475X - F110W) (pink) and (F475W - F850LP) (blue).}
    \label{fig:sim_GCs}
\end{figure}

We then created a set of simulated GCs using the Padova online tool (see \cite{bressan2012parsec}, \cite{chen2014parsec}, \cite{chen2015parsec}, \cite{tang2014parsec}, \cite{marigo2017parsec}, \cite{pastorelli2019parsec}, and \cite{pastorelli2020parsec} for details), specifying a fixed age of 12 Gyr and a total mass of $10^5$ $\textrm{M}_{\odot}$ for each cluster, and allowing the metallicity to increase incrementally from $[\textrm{M}/\textrm{H}] = -2.2$ to $[\textrm{M}/\textrm{H}] = 0.3$ in steps of 0.1 dex.  Using the Padova tool output, we plotted the resulting mock \textit{HST} observations in both ($\textrm{F}475\textrm{W} - \textrm{F}850\textrm{LP}$), the calibrated index, and ($\textrm{F}475\textrm{X} - \textrm{F}110\textrm{W}$), our index, versus the input metallicity.  Because our observations allow us to measure color, we inverted the results to obtain metallicity as a function of color.  The metallicity-color relations, fitted with exponential equations, appear in Figure \ref{fig:sim_GCs}.

\begin{figure}
    \figurenum{11}
    \centering
    \includegraphics[width=0.47\textwidth]{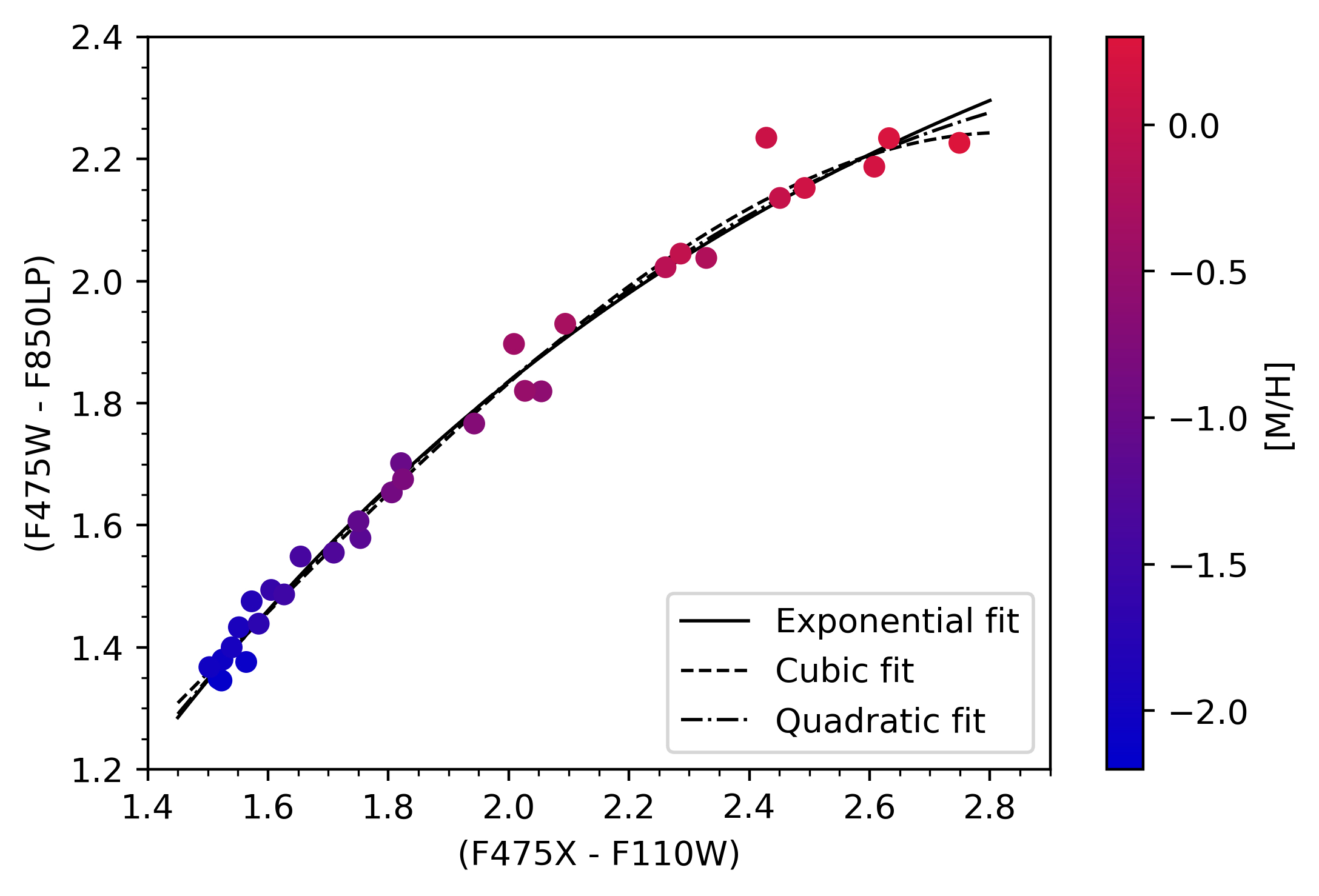}
    \caption{(F475W - F850LP) vs. (F475X - F110W) observations of simulated GCs.  Points are colored by metallicity.  We fitted exponential, cubic, and quadratic models, and found that the exponential model best captured the reddest GCs.}
    \label{fig:sim_GC_colour_colour}
\end{figure}

Our color-to-color-to-metallicity conversion is summarized in Figure \ref{fig:sim_GC_colour_colour}; it allowed us to express our observed ($\textrm{F}475\textrm{X} - \textrm{F}110\textrm{W}$) GC colors in terms of ($\textrm{F}475\textrm{W} - \textrm{F}850\textrm{LP}$), and then metallicity:

\begin{equation}
    (\textrm{F475W}-\textrm{F850LP}) = 2.785 - 5.000e^{-0.830(\rm F475X-F110W)}
\end{equation}

\begin{multline}
    \rm [M/H] = (4.456\cdot10^8(F475W-F850LP) \\ - 5.996\cdot10^8)^{1/2} / 6684 - 2.847
\end{multline}

\section{Modeling MDF shape} \label{sec:DGfits}

\subsection{Double Gaussian fits}

\begin{figure*}
    \figurenum{12}
    \centering
    \includegraphics[width=\textwidth]{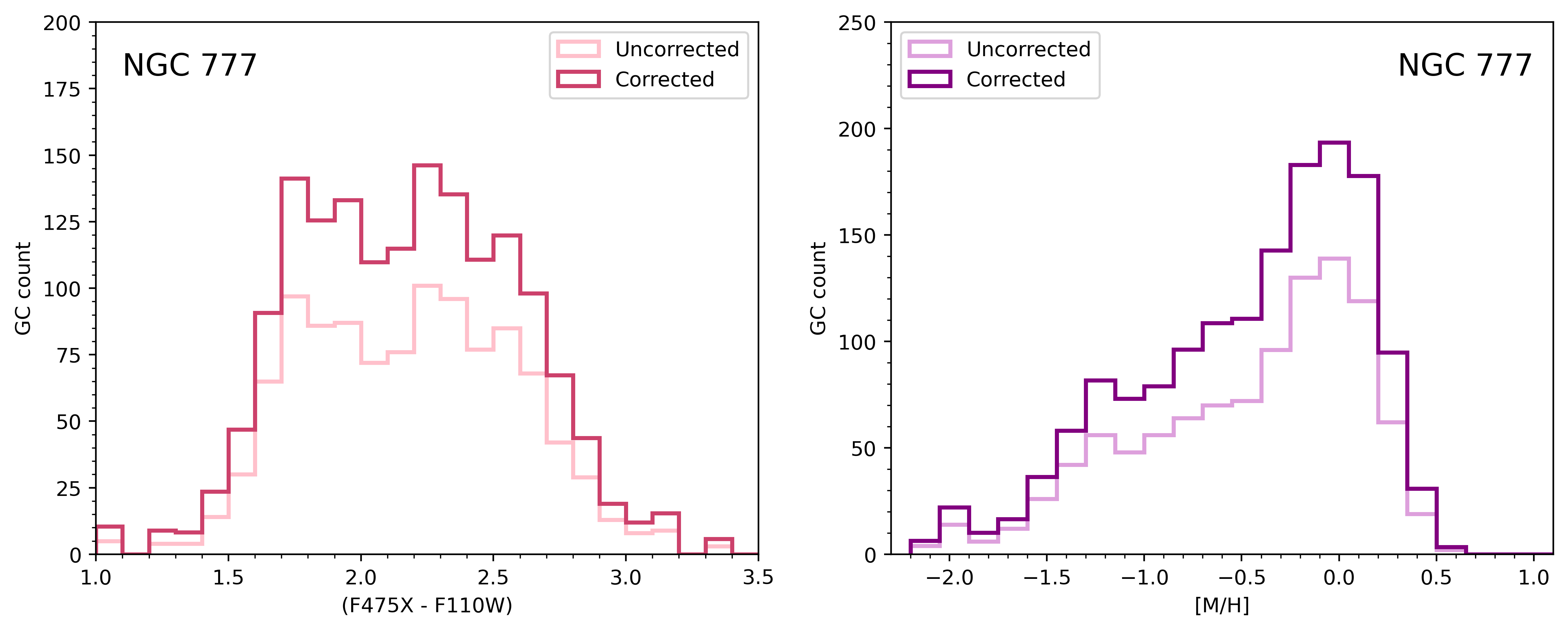}
    \caption{\textit{Left:} color histogram for NGC 777.  \textit{Right:} metallicity histogram for NGC 777.  Completeness-corrected GC counts are shown in darker colors.  The color-to-metallicity transformation described in Section \ref{sec:metallicity} compresses the red tail of the CDF and elongates the blue tail.  Completeness corrections boost GC counts while preserving the shape of both histograms.  (Note that the data in this figure have not undergone the final magnitude and radius cuts described in Section \ref{sec:completeness}; all data in subsequent figures have.)}
    \label{fig:ngc777_hists}
\end{figure*}

The completeness-sky brightness relation described in Section \ref{sec:completeness} and the color-metallicity conversion from Section \ref{sec:metallicity} enabled us to turn our directly observed color indices into more physically meaningful metallicity values.  Figure \ref{fig:ngc777_hists} shows a typical result of both processes: the completeness correction brought up the total GC numbers, though with little change to the shape of the CDF or MDF, while the color-to-metallicity conversion compresses the red GCs into a metal-rich peak and stretches the blue GCs into a metal-poor tail (note that the data in this figure have not undergone the final magnitude and radius cuts described in Section \ref{sec:completeness}; all data in subsequent figures have).  The differences, though small, between the raw and completeness-corrected MDFs are in large part due to the shallow radial metallicity gradients that are also an observed feature of these systems (see below): because the completeness corrections are a bit larger in the inner regions where the metal-richer clusters are a bit more prominent, the metal-richer side of the MDF is boosted slightly more.

\begin{table*}
    \caption{GMM statistics - original GC counts}
    \label{tab:GMM_check}
    \begin{tabular}{c|cccccccccc}
        \toprule
        Galaxy & $N_{GCs}$ & $\mu_{mp}$ & $\sigma_{mp}$ & $\mu_{mr}$ & $\sigma_{mr}$ & $\chi$ & $p$ & Dip & $p_{dip}$ \\
        \tableline
        NGC 57 & 1318 & -0.804 & 0.533 & -0.011 & 0.194 & 263.20 & $<0.0001$ & 1.98 & 0.90 \\
        NGC 410 & 810 & -0.807 & 0.591 & -0.057 & 0.193 & 88.73 & $<0.0001$ & 1.71 & 0.77 \\
        NGC 533 & 1666 & -0.854 & 0.541 & -0.047 & 0.195 & 189.93 & $<0.0001$ & 1.71 & 0.14 \\
        NGC 741 & 632 & -0.851 & 0.481 & -0.147 & 0.223 & 61.57 & $<0.0001$ & 1.88 & 0.92 \\
        NGC 777 & 1062 & -0.881 & 0.556 & -0.042 & 0.222 & 259.07 & $<0.0001$ & 1.98 & 0.99 \\
        NGC 1016 & 1135 & -0.828 & 0.550 & -0.054 & 0.224 & 178.09 & $<0.0001$ & 1.84 & 0.98 \\
        NGC 1129 & 2144 & -0.669 & 0.530 & -0.085 & 0.190 & 203.45 & $<0.0001$ & 1.47 & 0.99 \\
        NGC 1600 & 1526 & -0.851 & 0.557 & -0.075 & 0.193 & 256.85 & $<0.0001$ & 1.86 & 0.86 \\
        NGC 2340 & 1404 & -0.710 & 0.534 & -0.029 & 0.210 & 150.10 & $<0.0001$ & 1.68 & 0.86 \\
        NGC 3158 & 1491 & -1.027 & 0.663 & -0.151 & 0.314 & 145.68 & $<0.0001$ & 1.69 & 0.99 \\
        NGC 3842 & 1420 & -0.993 & 0.578 & -0.189 & 0.241 & 123.62 & $<0.0001$ & 1.82 & 0.46 \\
        NGC 4073 & 1836 & -0.898 & 0.557 & -0.079 & 0.209 & 350.77 & $<0.0001$ & 1.95 & 0.99 \\
        NGC 4839 & 1719 & -0.938 & 0.551 & -0.120 & 0.223 & 162.27 & $<0.0001$ & 1.95 & 0.96 \\
        NGC 4914 & 607 & -1.185 & 0.606 & -0.220 & 0.326 & 121.07 & $<0.0001$ & 1.98 & 0.99 \\
        NGC 7242 & 2178 & -0.633 & 0.536 & -0.004 & 0.203 & 185.23 & $<0.0001$ & 1.55 & 0.99 \\
         \tableline
    \end{tabular}
\end{table*}

To quantify the shape of the MDFs, we modeled each metallicity histogram with a double Gaussian curve:

\begin{equation}
    N_{GCs,comp} = A_{mp}e^{{\frac{-(x-\mu_{mp})^2}{2\sigma_{mp}^2}}} + A_{mr}e^{{\frac{-(x-\mu_{mr})^2}{2\sigma_{mr}^2}}}
    \label{eq:DG}
\end{equation} with each peak (metal-poor, subscript $mp$, and metal-rich, subscript $mr$) characterized by an amplitude $A$, a mean $\mu$, and a standard deviation $\sigma$.

Testing on the uncorrected GC counts with the Gaussian mixture modeling program from \cite{muratov2010gmm} strongly favored a double Gaussian over a single Gaussian model, and showed no difference between double Gaussian and multi-Gaussian models (that is, any extra modes above $n=2$ were damped down to zero in the GMM solution).  Table \ref{tab:GMM_check} shows GMM test results from the original (i.e. not completeness-corrected) GC metallicity distributions; the $\chi$ and $p$ values are based on a null hypothesis of a unimodal Gaussian distribution, and $D$ is a measure of peak separation.  For details on $D$, see Section \ref{sec:environment} and Equation \ref{eq:sepD}.

\begin{table*}
    \centering
    \caption{Double Gaussian fitting}
    \label{tab:DGfit}
    \small
    \begin{tabular}{>{\centering\arraybackslash}p{0.09\textwidth}|>{\centering\arraybackslash}p{0.1\textwidth}>{\centering\arraybackslash}p{0.1\textwidth}>{\centering\arraybackslash}p{0.1\textwidth}>{\centering\arraybackslash}p{0.1\textwidth}>{\centering\arraybackslash}p{0.1\textwidth}>{\centering\arraybackslash}p{0.1\textwidth}>{\centering\arraybackslash}p{0.05\textwidth}>{\centering\arraybackslash}p{0.05\textwidth}} 
        \toprule
        Galaxy & $A_{mp}$ (GCs) & $\mu_{mp}$ [M/H] & $\sigma_{mp}$ [M/H] & $A_{mr}$ (GCs) & $\mu_{mr}$ [M/H] & $\mu_{mr}$ [M/H] & $D$ & $f_b$ \\
        \tableline
        NGC 57 & 68.6 $\pm$ 4.6 & -0.73 $\pm$ 0.10 & 0.51 $\pm$ 0.08 & 161.9 $\pm$ 14.2 & 0.02 $\pm$ 0.01 & 0.20 $\pm$ 0.02 & 1.947 & 0.248 \\
        NGC 410 & 48.3 $\pm$ 5.0 & -0.63 $\pm$ 0.11 & 0.55 $\pm$ 0.09 & 65.2 $\pm$ 12.4 & 0.06 $\pm$ 0.02 & 0.13 $\pm$ 0.03 & 1.711 & 0.334 \\
        NGC 533 & 106.1 $\pm$ 6.2 & -0.59 $\pm$ 0.06 & 0.56 $\pm$ 0.05 & 103.8 $\pm$ 14.9 & 0.04 $\pm$ 0.02 & 0.13 $\pm$ 0.02 & 1.545 & 0.407 \\
        NGC 741 & 32.4 $\pm$ 4.8 & -0.97 $\pm$ 0.09 & 0.31 $\pm$ 0.09 & 57.2 $\pm$ 5.9 & -0.18 $\pm$ 0.05 & 0.24 $\pm$ 0.04 & 2.828 & 0.192 \\
        NGC 777 & 50.8 $\pm$ 3.3 & -0.72 $\pm$ 0.11 & 0.52 $\pm$ 0.08 & 122.5 $\pm$ 11.4 & 0.00 $\pm$ 0.01 & 0.23 $\pm$ 0.02 & 1.803 & 0.216 \\
        NGC 1016 & 38.9 $\pm$ 3.8 & -0.61 $\pm$ 0.14 & 0.53 $\pm$ 0.10 & 59.5 $\pm$ 11.4 & 0.00 $\pm$ 0.01 & 0.20 $\pm$ 0.04 & 1.678 & 0.301 \\
        NGC 1129 & 128.6 $\pm$ 9.3 & -0.66 $\pm$ 0.10 & 0.47 $\pm$ 0.07 & 161.0 $\pm$ 29.1 & -0.06 $\pm$ 0.02 & 0.20 $\pm$ 0.03 & 1.671 & 0.336 \\
        NGC 1600 & 97.4 $\pm$ 7.6 & -0.59 $\pm$ 0.08 & 0.53 $\pm$ 0.05 & 124.4 $\pm$ 16.5 & -0.03 $\pm$ 0.02 & 0.19 $\pm$ 0.03 & 1.400 & 0.325 \\
        NGC 2340 & 52.3 $\pm$ 6.2 & -0.66 $\pm$ 0.19 & 0.48 $\pm$ 0.13 & 74.3 $\pm$ 21.2 & 0.02 $\pm$ 0.03 & 0.23 $\pm$ 0.05 & 1.780 & 0.297 \\
        NGC 3158 & 31.0 $\pm$ 4.2 & -0.78 $\pm$ 0.21 & 0.58 $\pm$ 0.13 & 47.9 $\pm$ 12.1 & -0.06 $\pm$ 0.02 & 0.27 $\pm$ 0.05 & 1.581 & 0.278 \\
        NGC 3842 & 42.9 $\pm$ 8.0 & -0.87 $\pm$ 0.31 & 0.55 $\pm$ 0.19 & 54.6 $\pm$ 25.3 & -0.14 $\pm$ 0.05 & 0.27 $\pm$ 0.08 & 1.675 & 0.302 \\
        NGC 4073 & 47.5 $\pm$ 6.2 & -0.70 $\pm$ 0.16 & 0.58 $\pm$ 0.10 & 114.5 $\pm$ 14.1 & -0.06 $\pm$ 0.02 & 0.21 $\pm$ 0.03 & 1.451 & 0.297 \\
        NGC 4839 & 49.9 $\pm$ 2.7 & -0.78 $\pm$ 0.07 & 0.50 $\pm$ 0.06 & 64.0 $\pm$ 7.8 & -0.05 $\pm$ 0.02 & 0.18 $\pm$ 0.02 & 1.944 & 0.348 \\
        NGC 4914 & 18.3 $\pm$ 4.0 & -0.48 $\pm$ 0.22 & 0.47 $\pm$ 0.12 & 38.8 $\pm$ 9.3 & 0.04 $\pm$ 0.02 & 0.20 $\pm$ 0.04 & 1.434 & 0.252 \\
        NGC 7242 & 153.9 $\pm$ 7.5 & -0.41 $\pm$ 0.04 & 0.46 $\pm$ 0.03 & 90.8 $\pm$ 16.2 & 0.08 $\pm$ 0.02 & 0.14 $\pm$ 0.03 & 1.419 & 0.459 \\
        \tableline
    \end{tabular}
\end{table*}

GMM, the most widely used fitting code for GCS studies, takes into account previous research into the nature of GCS metallicity distributions and builds on the copious evidence for metal-rich and metal-poor GC subpopulations in many galaxies \citep[e.g.][]{zepf1993bimodal,forbes1997bimodal,kundu2001bimodal,larsen2001,brodie2006colour,peng2006nonlinear,arnold2011bimodal,brodie2012colour,kim2013bimodal,cantiello2014bimodal,brodie2014bimodal}, so the question now is: how do those subpopulations combine to form the full-GCS distributions that we observe?

A more basic question regarding uni- vs.\ bi-modality can be posed by applying a dip test to our results for the MDFs \citep{hartigan1985dip,gebhardt1999dip}, with numerical results as listed in the last two columns of Table \ref{tab:GMM_check}.  In most cases the probability $p$(dip) indicates that the MDF is best considered to be unimodal ($p > 0.9$), with only NGC 533 and NGC 3842 strongly favoring multimodality.  However, the dip test is only useful when the underlying form of the distribution is unknown.  While most of our MDFs are visibly unimodal (that is, they have only one clear peak), it is simultaneously true that they are all skewed and the GMM fitting strongly rejects any fit with a unimodal Gaussian.  The dip test was last used for GCSs in \cite{gebhardt1999dip}, and since then a double Gaussian mixture has in most cases been found to be an appropriate empirical model for both GCS color and metallicity distributions.

\begin{figure}
    \figurenum{13}
    \centering
    \includegraphics[width=0.47\textwidth]{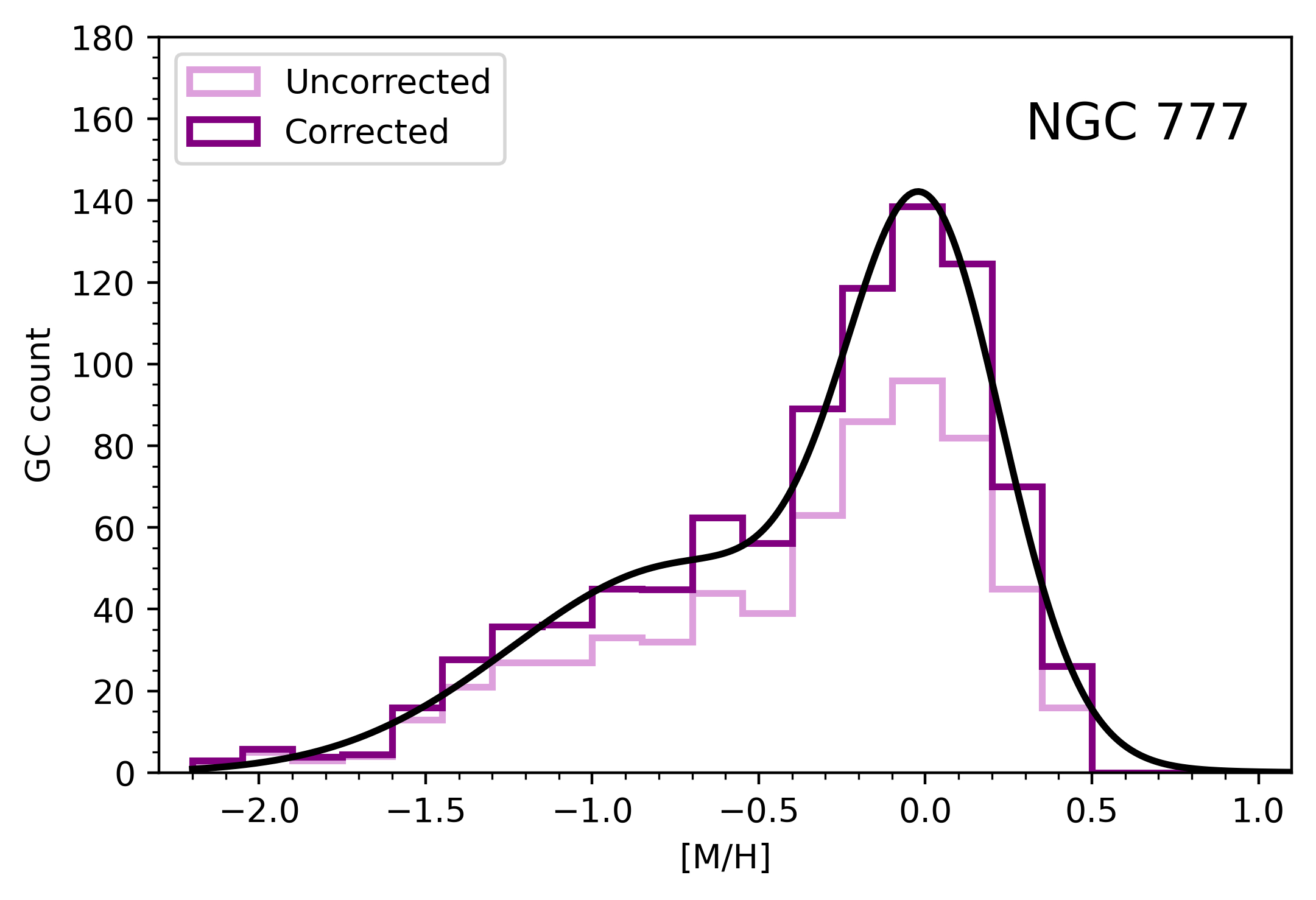}
    \caption{The fitted double Gaussian curve (in black) for NGC 777.  This galaxy is a typical example, with two visible subpopulations, a tall and narrow metal-rich peak, and a shallower and wider metal-poor tail.}
    \label{fig:ngc777_DGfit}
\end{figure}

\begin{figure*}
    \figurenum{14}
    \centering
    \includegraphics[width=\textwidth]{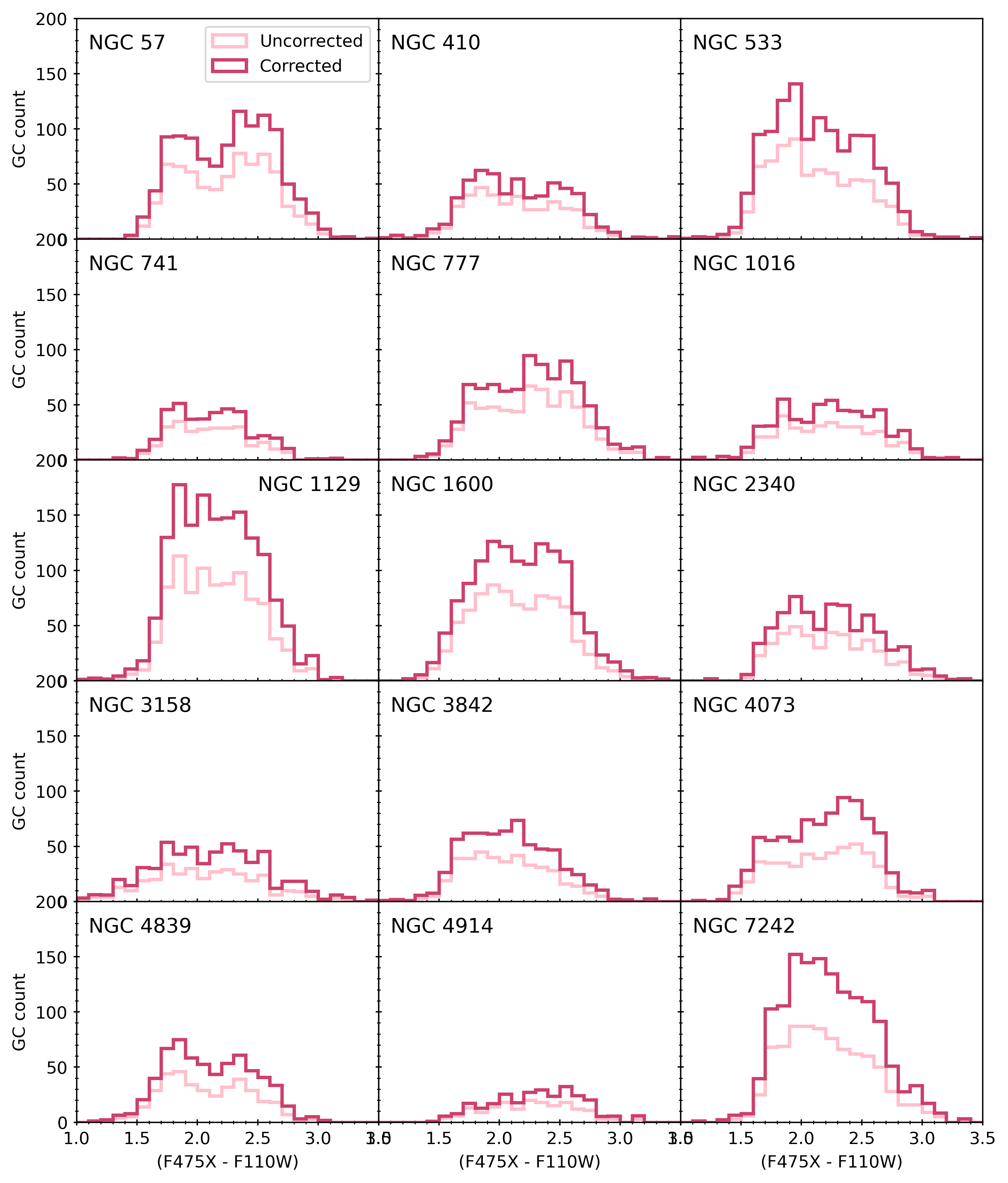}
    \caption{Color histograms for all 15 galaxies in the sample.  These are the final GC samples after all culling procedures (as opposed to the example color histogram in Figure \ref{fig:ngc777_hists}, which has not been fully culled).}
    \label{fig:all_colour_hists}
\end{figure*}

\begin{figure*}
    \figurenum{15}
    \centering
    \includegraphics[width=\textwidth]{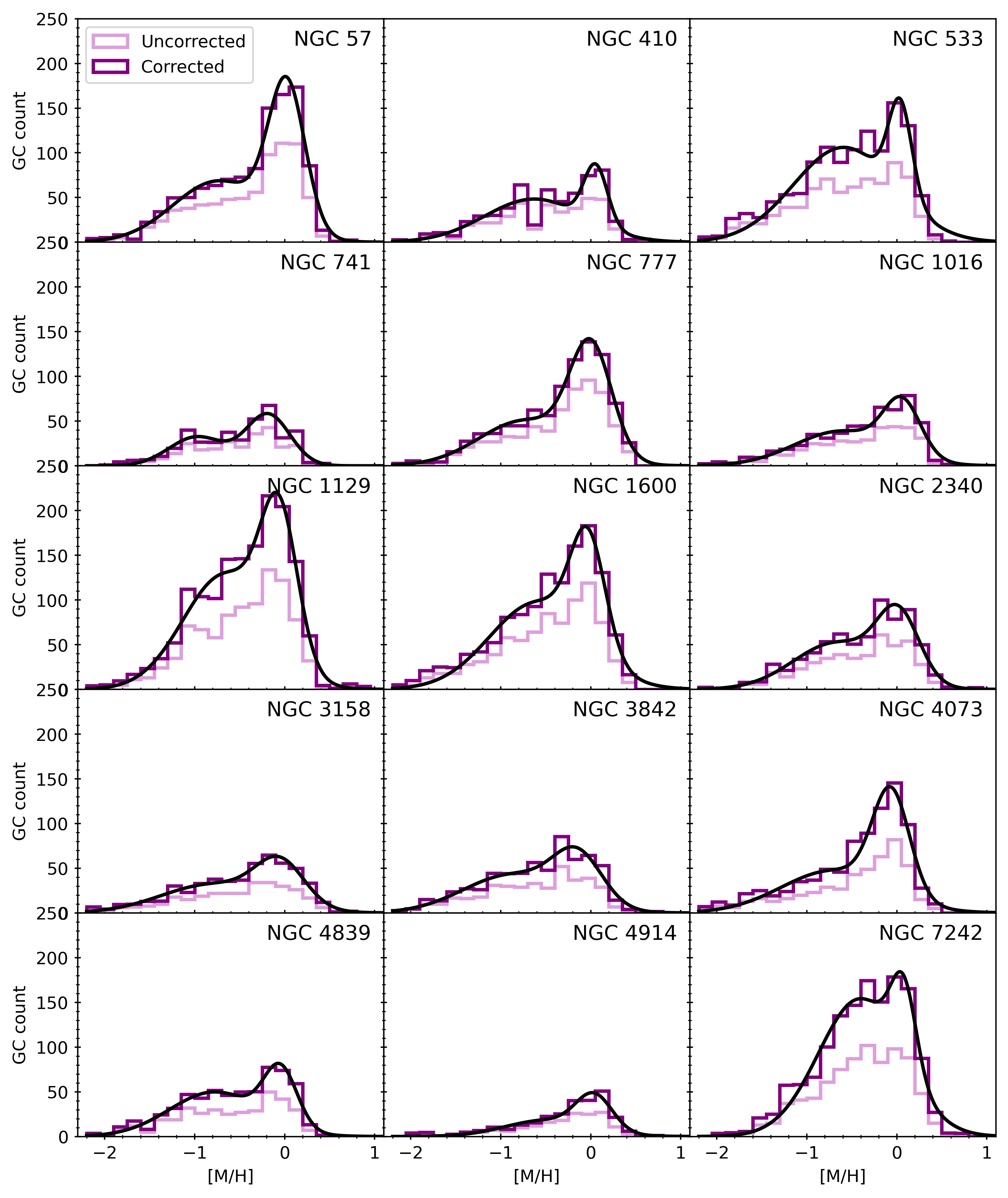}
    \caption{Fitted double Gaussian curves for all 15 galaxies in the sample.  There are wide ranges in our sample of total GC counts, metal-poor peak amplitude, distance between peaks, etc., but all GCSs are well modeled with a double Gaussian.}
    \label{fig:all_hists}
\end{figure*}

Figure \ref{fig:ngc777_DGfit} shows the NGC 777 double Gaussian model as an example, and Figure \ref{fig:all_hists} compares the MDFs and double Gaussian models for all fifteen galaxies in our sample.  All the double Gaussian fit parameters can be found in Table \ref{tab:DGfit}.  Figures \ref{fig:all_colour_hists} and \ref{fig:all_hists} can be compared as in Figure \ref{fig:ngc777_hists}.

\subsection{Metallicity gradients}

\begin{figure*}
    \figurenum{16}
    \includegraphics[width=0.99\textwidth]{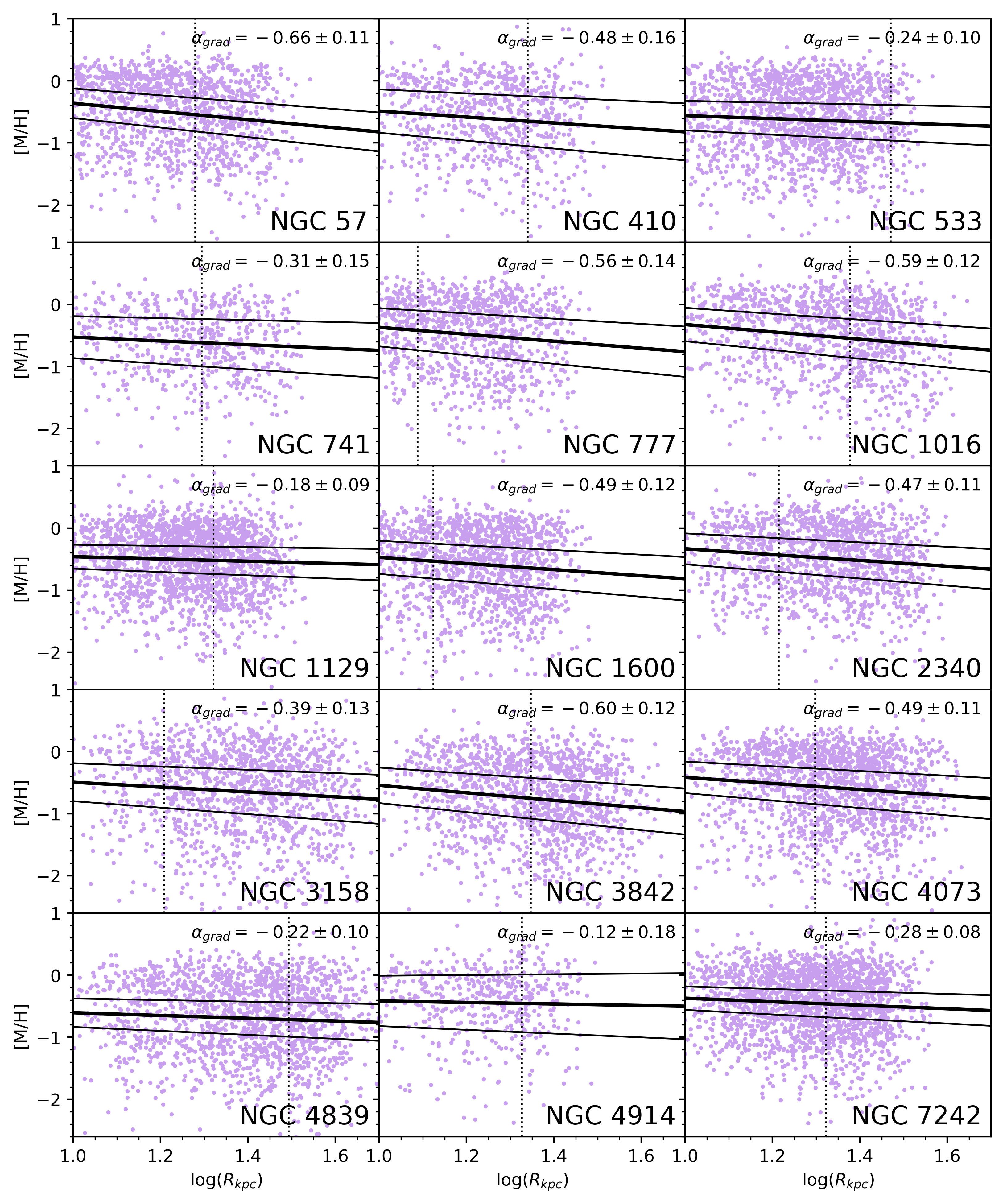}
    \caption{Metallicity vs. projected radius for each galaxy's GCS.  Thick and thin solid lines show the fitted metallicity gradient and one-sigma errors, respectively; vertical dotted lines denote $2R_{e}$ (see Table \ref{tab:basics}).  Slopes and one-sigma uncertainties} are given in the top right corner of each plot.  Metallicity gradients for our galaxy sample are in line with literature gradients.
    \label{fig:metgrad}
\end{figure*}

In addition to testing whether the standard double Gaussian was an appropriate model for the GCSs in our sample, we modeled the metallicity gradients of our galaxies and checked them against comparable results from the literature.  Figure \ref{fig:metgrad} shows metallicity versus projected radius, along with one-sigma uncertainties.  Slopes, uncertainties, and significance compared to a flat metallicity gradient can also be found in Table \ref{tab:metgrad}.

\begin{table}
    \centering
    \caption{Metallicity gradients}
    \label{tab:metgrad}
    \begin{tabular}{>{\centering\arraybackslash}p{0.09\textwidth}|>{\centering\arraybackslash}p{0.06\textwidth}>{\centering\arraybackslash}p{0.11\textwidth}>{\centering\arraybackslash}p{0.12\textwidth}}
        \toprule
        Galaxy & $\alpha_{grad}$ & Uncertainty & Significance \\
        \tableline
        NGC 57 & -0.66 & $\pm$0.11 & $6.0\sigma$ \\
        NGC 410 & -0.48 & $\pm$0.16 & $3.0\sigma$ \\
        NGC 533 & -0.24 & $\pm$0.10 & $2.4\sigma$ \\
        NGC 741 & -0.31 & $\pm$0.15 & $2.1\sigma$ \\
        NGC 777 & -0.56 & $\pm$0.14 & $4.0\sigma$ \\
        NGC 1016 & -0.59 & $\pm$0.12 & $4.9\sigma$ \\
        NGC 1129 & -0.18 & $\pm$0.09 & $2.0\sigma$ \\
        NGC 1600 & -0.49 & $\pm$0.12 & $4.1\sigma$ \\
        NGC 2340 & -0.47 & $\pm$0.11 & $4.3\sigma$ \\
        NGC 3158 & -0.39 & $\pm$0.13 & $3.0\sigma$ \\
        NGC 3842 & -0.60 & $\pm$0.12 & $5.0\sigma$ \\
        NGC 4073 & -0.49 & $\pm$0.11 & $4.5\sigma$ \\
        NGC 4839 & -0.22 & $\pm$0.10 & $2.2\sigma$ \\
        NGC 4914 & -0.12 & $\pm$0.18 & $0.7\sigma$ \\
        NGC 7242 & -0.28 & $\pm$0.08 & $3.5\sigma$ \\
         \tableline
    \end{tabular}
\end{table}

Fitting a power law to the metallicity gradients of the form [M/H] $= const + \alpha_{grad} {\rm log} R$, we found an average slope of $\alpha_{grad} = -0.41$, with an rms scatter of $\pm 0.13$.  These values are consistent with the results for numerous other galaxies over a wide range of luminosities and appear to be a common feature of GCSs:  a shallow radial decrease in mean metallicity, though with galaxy-to-galaxy scatter of $0.1-0.2$ dex 
\citep[e.g.][]{geisler1996metgrad,harris2009metgrad,liu2011metgrad,forbes2018metgrad,harris2023}.

Despite the shallow mean gradients, the mean of the metal-rich GC subpopulation ($\mu_{mr}$ in particular) is very similar from one galaxy to another, hovering around [M/H] $\sim 0$ with a scatter of only a few percent.  A shallow trend for $\mu_{mr}$ to increase with galaxy mass roughly as $Z \sim M_*^{0.2}$ has been established in previous surveys \citep[e.g.][]{larsen2001,strader2004,peng2006nonlinear}, though our selection of targets with nearly the same stellar masses would prevent that trend from showing up here.  The mean absolute value of $\mu_{mr} \sim 0.0$ is, however, more metal-rich by a few tenths of a dex than has been found for red GCs in spectroscopic surveys of massive galaxies (cf. the references cited above), and also from photometric surveys based on optical color indices \citep[e.g.][]{peng2006nonlinear,harris2023}.  As described above, the metallicity values we obtain are the direct result of one particular set of stellar model transformations from the F475X and F110W filters; these need to be investigated further with additional sets of models and new empirical transformations to optical indices.

\section{Comparison with host galaxy environment} \label{sec:environment}

Because the amplitudes from our double Gaussian fits are influenced by the number of GCs in the sample for each galaxy (and will be divided by GC count for the rest of this analysis) and by the overall shape of the curve, the key parameters are the metal-poor and metal-rich modes $\mu_{mp}$ and $\mu_{mr}$ and the metal-poor and metal-rich widths (standard deviations) $\sigma_{mp}$ and $\sigma_{mr}$.  The metal-rich peaks are quite uniform throughout our galaxy sample, with a mode of $\mu_{mr} \sim 0.0$ and width ranging from $0.1 \lesssim \sigma_{mr} \lesssim 0.3$.  Most differences between GCSs arise with the metal-poor peak, which can lie anywhere from $\sim 0.5$ to $\sim 0.8$ dex away from the metal-rich peak.

In addition to comparing the double Gaussian parameters themselves, we calculated the mean metallicity for each GCS, the difference between modes ($\mu_{mr} - \mu_{mp}$), a second peak separation metric taking into account peak width:
\begin{equation}
    D = \frac{|\mu_{mp} - \mu_{mr}|}{\sqrt{\frac{\sigma_{mp}^2 + \sigma_{mr}^2}{2}}}
    \label{eq:sepD}
\end{equation}
\citep{muratov2010gmm}, and the blue fraction $f_{b}$, recovered from the double Gaussian solution.

In general, very few GCS variables showed any signs of being correlated with environmental metrics---see Table \ref{tab:spearman} for significant Spearman coefficients---with the exception of the blue fraction and the normalized metal-rich amplitude, which are themselves related.  Because blue fraction is a more physically relevant parameter, our analysis will focus on it rather than on normalized metal-rich amplitude.

\begin{figure*}
    \figurenum{17}
    \centering
    \includegraphics[width=\textwidth]{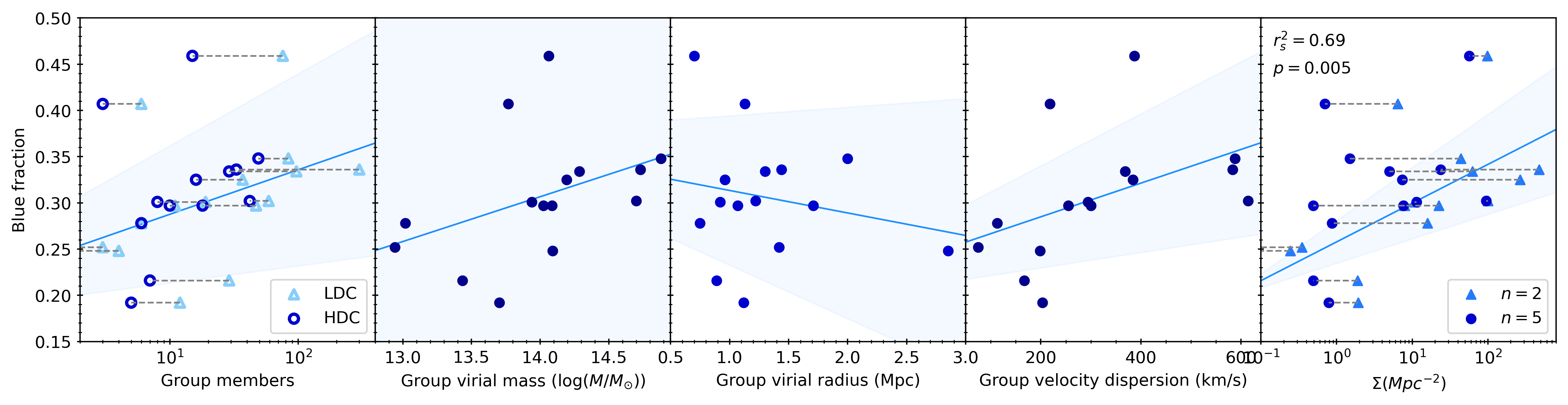}
    \caption{From left to right: GC blue fraction vs. number of group members, group virial mass, group virial radius, group velocity dispersion, and \textit{n}th-nearest neighbor surface density.  In the leftmost column, open circles indicate HDC measurements, and open triangles LDC measurements.  In the rightmost column, circles indicate $n=5$ figures, and triangles indicate $n=2$ figures.  Dashed lines connect points for the same galaxy.  Blue lines and shading denote linear fits with one-sigma intervals.  In the rightmost panel, the Spearman coefficient, p-value, and linear fit are for $n=2$.}
    \label{fig:Bfrac_enviro}
\end{figure*}

Figure \ref{fig:Bfrac_enviro} shows $f_b$ versus all environmental metrics derived from \cite{crook2007groups}, with fitted linear models and one-sigma range shaded.  Functions of group member count, group virial mass, group virial radius, and group velocity dispersion all produced large uncertainty for linear fit parameters and visual inconsistency when plotted.  In contrast, \textit{n}th-nearest neighbor surface density produces a smaller linear fit uncertainty and a more visually consistent positive trend (i.e. galaxies in denser neighborhoods tend to have a higher blue fraction), in addition to a significant Spearman coefficient.  The relation is weak, but stronger than the poorly constrained results for other variable combinations.  The linear fit parameters for $f_b$ versus $\Sigma_2$ are:

\begin{equation}
    \langle f_b\rangle = 0.042\log(\Sigma_2) + 0.258
    \label{eq:Bfrac_hood_lin}
\end{equation}

\begin{table}
    \centering
    \caption{Significant GCS-environment comparisons}
    \label{tab:spearman}
    \scriptsize
    \begin{tabular}{c|c|cc}
        \toprule
          \shortstack{GCS \\ property} & \shortstack{Environmental \\ property} & $r^2_s$ & p-value \\
        \tableline
        Blue fraction & LDC members & 0.59 & 0.02 \\
        Blue fraction & HDC members & 0.56 & 0.03 \\
        Blue fraction & Group virial mass & 0.56 & 0.03 \\
        Blue fraction & Group velocity dispersion & 0.72 & 0.002 \\
        Blue fraction & $\Sigma_2$ & 0.69 & 0.005 \\
        Blue fraction & $\Sigma_5$ & 0.59 & 0.02 \\
        Metal-rich amplitude & $\Sigma_5$ & -0.53 & 0.04 \\
        \tableline
    \end{tabular}
\end{table}

\begin{figure}
    \figurenum{18}
    \centering
    \includegraphics[width=0.5\textwidth]{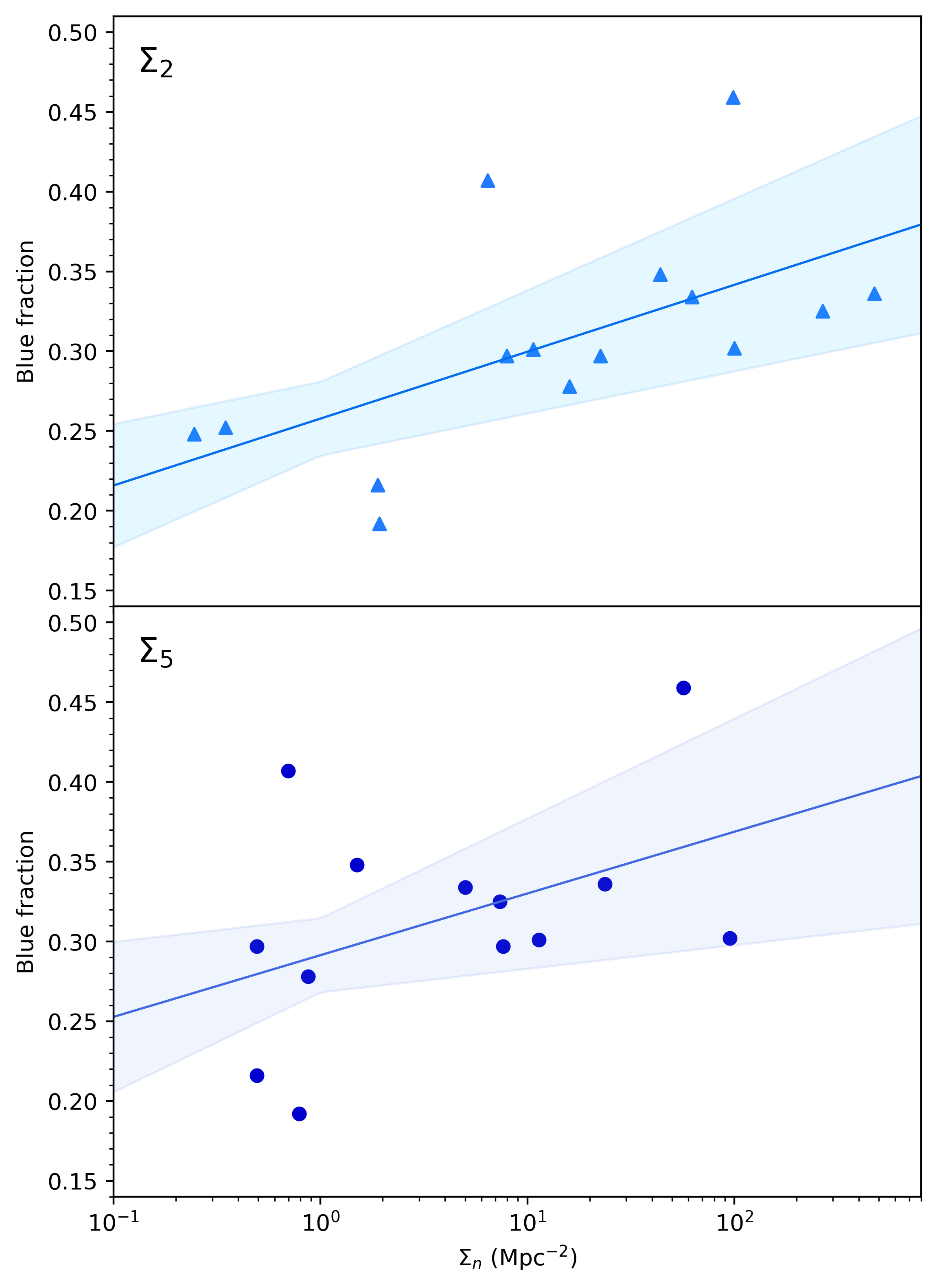}
    \caption{Blue fraction vs. \textit{n}th-nearest neighbor surface density with a linear fit (line) and one-sigma uncertainty (shaded area).  \textit{Top:} $n=2$.  \textit{Bottom:} $n=5$.}
    \label{fig:Bfrac_hood_fit}
\end{figure}

The correlation we find, i.e. that increased GC blue fraction is associated with high local number densities of satellite galaxies, is at least superficially consistent with expectations from current theory that a high fraction of the metal-poor GCs in BCGs have been accreted from nearby satellites (see Section \ref{sec:intro}).  It is not yet clear, however, whether the shallowness of this correlation and the others shown in Figure \ref{fig:Bfrac_enviro} is in quantitative agreement with present models.  The present observational work is necessarily focused on the inner to mid-halo region because of the restricted field of view of the cameras, whereas the theoretical simulations tend to provide global correlations.

\section{Discussion and Summary} \label{sec:summary}

In this study, we used \textit{HST} images of fifteen massive BCGs selected from the MASSIVE survey to investigate the relationship between GCS metallicity and host galaxy environment.  We created a model of completeness across the varying background brightness in our images, and a model for GC metallicity as a function of the \textit{HST} color index ($\textrm{F}475\textrm{X} - \textrm{F}110\textrm{W}$).  After fitting a double Gaussian curve to each MDF, we compared the GCS metallicity parameters to environmental metrics derived from \cite{crook2007groups} and found a weak but consistent correlation between the GCS blue fraction and \textit{n}th-nearest neighbor density, but no statistically significant relationship between any other variables.

It should be noted that we consider this work as a pilot study into the relationship between GCS metallicity distribution and host galaxy environment.  There are several issues that need further exploration:

\begin{enumerate}
    \item The conversion from color to metallicity used here is a preliminary step.  We plan to compare the Padova models to other simple stellar population (SSP) models, to build in other features such as age/metallicity relations, and to develop direct empirical transformations between color indices from upcoming \textit{HST} observations.
    \item It is possible that the \textit{HST} images used in this work did not capture enough of our galaxies' halos for us to detect a strong relation between GCS and environmental parameters.  Recent observational studies such as \cite{hughes2022outerhalo} show that accreted GCs may remain far from their new host galaxy's center, well outside the WFC3 field of view at the distance of our galaxy sample.  To understand the full GCSs of our galaxies, we would need either a wider field of view or mosaic images.
    \item We should also consider the evolutionary clock of these galaxies---we are looking at them at a certain stage in their growth. The galaxies in sparser environments have essentially finished their hierarchical growth as they have no more satellites to absorb, while the ones in richer areas are still moving along their merger trees and will accrete more metal-poor GCs and increase their blue fractions in the future.  Environmentally driven differences may thus emerge more strongly as BCGs evolve beyond the present day.
    \item We need to consider how much host galaxy environment we include.  A galaxy's wider environment and more distant satellites may not have an appreciable effect on the inner halo; at what level of locality (or nonlocality) does environment begin to correlate more obviously with GCS properties?
\end{enumerate}

\section{Acknowledgements}


We acknowledge financial support from the Natural Sciences and Engineering Research Council of Canada (NSERC) through a Discovery Grant to WEH.  KH thanks Alison Sills, Laura Parker, Veronika Dornan, Claude Cournoyer-Cloutier, and Jerermy Karam for helpful discussions.

This research has made use of the SIMBAD database and the Vizier catalogue access tool, operated at CDS, Strasbourg, France \citep{wenger2000simbad}.


%

\facilities{\textit{HST} (WFC3)}


\software{Python (\href{https://www.python.org}{https://www.python.org}), NumPy \citep{harris2020numpy}, pandas \citep{team2020pandas}, Matplotlib \citep{hunter2007matplotlib}, pyraf \citep{stsci2012pyraf}, DOLPHOT \citep{dolphin2000dolphot}, ds9 \citep{smithsonian2000ds9}}



\bibliography{Bibliography.bib}{}
\bibliographystyle{aasjournal}



\end{document}